\documentclass[a4paper,11pt]{article}
\pdfoutput=1
\usepackage{jheppub}
\usepackage[T1]{fontenc}
\usepackage{xcolor}
\usepackage{booktabs}
\usepackage{multirow}

\newcommand{\qb}{\bar{q}}
\newcommand{\qp}{q'}
\newcommand{\qbp}{\bar{q}'}
\newcommand\id[1]{#1^{\rm id.}}
\newcommand{\XTO}[2]{\mathcal{#1}_3^{1,{\rm id.}#2}}
\newcommand{\XFZ}[2]{\mathcal{#1}_4^{0,{\rm id.}#2}}
\newcommand{\XTZ}[2]{\mathcal{#1}_3^{0,{\rm id.}#2}}
\newcommand{\XTZd}[3]{\mathcal{#1}_{3,#3}^{0,{\rm id.}#2}}
\newcommand{\MF}{{\rm MF}}
\newcommand{\as}{\alpha_s}
\newcommand{\e}{\epsilon}
\newcommand{\B}{\rm B}
\newcommand{\rto}{\leftarrow}
\newcommand{\MFKB}[2]{\mathbf{\Gamma}^{(#1)}_{#2}}
\newcommand{\MFK}[2]{\Gamma^{(#1)}_{#2}}
\newcommand{\Dn}[1]{\mathcal{D}_{#1}}
\newcommand{\der}{\mathrm{d}}
\def\hsig{\hat{\sigma}}
\def\LO{{\rm LO}}
\def\NLO{{\rm NLO}}
\def\NNLO{{\rm NNLO}}
\def\R{{\rm R}}
\def\RR{{\rm RR}}
\def\RV{{\rm RV}}
\def\VV{{\rm VV}}
\def\V{{\rm V}}
\def\S{{\rm S}}

\def\T{{\rm T}}
\def\U{{\rm U}}
\newcommand\epem{e^+e^-}
\def\ff{\hat{\mathcal{F}}}
\def\ft{\hat{\mathcal{T}}}
\def\q{\mathcal{Q}}
\def\NS{{\rm NS}}
\def\PS{{\rm PS}}

\newcommand{\NF}{N_F}
\newcommand{\conv}{\otimes}
\def\JET{J}

\title{Antenna subtraction at NNLO with identified hadrons}
\author{Thomas Gehrmann,}
\author{Giovanni Stagnitto}
\affiliation{Physik-Institut, Universit\"at Z\"urich,
  Winterthurerstrasse 190, CH-8057 Z\"urich, Switzerland}
\emailAdd{thomas.gehrmann@uzh.ch}
\emailAdd{giovanni.stagnitto@physik.uzh.ch}


\preprint{ZU-TH 39/22}

\abstract{  We extend the antenna
  subtraction method to include hadron fragmentation processes up to
  next-to-next-to-leading order (NNLO) in QCD in $e^+e^-$ collisions. To 
  handle collinear singularities associated with the fragmentation process, 
  we introduce fragmentation antenna functions in final-final
  kinematics with associated phase space mappings. These antenna functions are
  integrated over the relevant phase spaces, retaining their dependence on the 
  momentum fraction of the fragmenting parton. The integrated antenna functions are
  cross-checked against the known NNLO coefficient functions for identified
  hadron production from $\gamma^*/Z^* \to q\bar{q}$ and $H \to gg$ processes. }

\begin{document} 
\maketitle
\flushbottom

\section{Introduction}
\label{sec:intro}

The production of identified hadrons in high-energy particle collisions has been
among the very first observables studied in experimental particle physics. These
measurements were usually performed in a single-particle inclusive manner,
i.e.\ differential in the kinematics of the identified hadron but fully
inclusive over all hadronic activity in the event. Very extensive measurements
of single-inclusive hadron production were performed at $e^+e^-$
colliders~\cite{Lafferty:1995jt}. The resulting legacy data sets provide
important information on the transition from partons to hadrons and are used to
tune the parameters of empirical hadronisation
models~\cite{Andersson:1983ia,Webber:1986mc} that form the basis of modern Monte
Carlo event simulation.

In QCD, single-inclusive hadron production can be described by convoluting
single-inclusive parton production, which is calculable in perturbation theory,
with fragmentation functions (FF,~\cite{Field:1976ve,Field:1977fa}) that
parametrise the parton-to-hadron transition as a function of the fractional
momentum transfer.  The factorisation behaviour of FFs closely resembles that of
parton distribution functions (PDF), and they also fulfil Altarelli-Parisi
evolution equations~\cite{Altarelli:1977zs} in their associated resolution
scale. In contrast to a description based on empirical hadronisation models, the
FF framework can be systematically expanded~\cite{Altarelli:1979kv} in
perturbative QCD by computing higher-order corrections to the partonic
coefficient functions and the FF evolution kernels.  The coefficient functions
for single-inclusive hadron production are known to next-to-leading order (NLO)
for electron-proton~\cite{Altarelli:1979kv,Baier:1979sp,deFlorian:1997zj} and
proton-proton~\cite{Aversa:1988vb} collisions and to next-to-next-to-leading
order (NNLO) for electron-positron
annihilation~\cite{Rijken:1996ns,Mitov:2006ic}. The FF evolution kernels are
also known to NLO~\cite{Furmanski:1980cm} and NNLO~\cite{Almasy:2011eq}. The
initial conditions to the FF evolution equations reflect the non-perturbative
dynamics of the parton-to-hadron transition. They can not be computed from first
principles in perturbative QCD, and are typically determined by global fits to
experimental data on single-inclusive hadron cross
sections~\cite{Albino:2008gy,Metz:2016swz,Albino:2008fy,deFlorian:2014xna,Sato:2016wqj,Anderle:2015lqa,Bertone:2017tyb,Borsa:2022vvp,Khalek:2022vgy}
for various hadron species.

The fragmentation function formalism was initially developed for light quarks,
where the initial conditions to the evolution equations are purely
non-perturbative. In the case of heavy quarks, the quark mass acts as an
infrared regulator that prevents exactly collinear emissions in the
fragmentation process (dead-cone
effect,~\cite{Dokshitzer:1991fd,ALICE:2021aqk}). By introducing a perturbative
contribution to the heavy-quark fragmentation functions~\cite{Mele:1990cw}, it
is possible to incorporate the heavy quark mass effects for observables
including identified heavy hadrons (or identified heavy quark
jets,~\cite{Bauer:2013bza}) into otherwise fully massless calculations at higher
orders, and resum large logarithms of collinear origin~\cite{Cacciari:1998it}.
The perturbative heavy-quark FFs were computed up to
NNLO~\cite{Mele:1990yq,Melnikov:2004bm} and augmented by soft-gluon
resummation~\cite{Cacciari:2001cw}.  They were used especially in NLO
calculations of identified heavy hadron production.

Identified hadrons play an increasingly important role in precision measurements
at the LHC, for example in cross sections in association with a vector boson or
a photon, where they are relevant to determining the flavour decomposition of
PDFs, or in the study of hadron production in jet substructure
studies~\cite{Chang:2013rca,Larkoski:2017jix,Li:2021zcf} or hadron-in-jet
production~\cite{OPAL:2004prv,Kaufmann:2015hma,Anderle:2017cgl}.

These collider measurements do however not fall into the class of
single-inclusive hadron production observables, since their final state
definition involves a set of criteria that is applied not only to the identified
hadron momentum but to the full final state of the event. These criteria can be
fiducial cuts on other particles, or more generally any type of infrared-safe
event classification criterion, such as the application of a jet algorithm or of
an event shape requirement. While FFs for the hadron species under consideration
may well be known from a global fit, their application to these less inclusive
observables is prevented by the incomplete understanding of the parton-level
cross sections for identified hadrons in the presence of generic event-based
fiducial selection criteria. Instead, data on these processes is typically
compared only to Monte Carlo event simulation using empirical hadronisation
models, which offer a considerably larger degree of flexibility in adapting to
specific final state definitions than the FF framework, however with the
drawback of considerably lower theory precision.

Higher order corrections to single-inclusive coefficient functions are typically
obtained by analytical integration of the relevant parton-level subprocesses
from real and virtual contributions. For fully exclusive fiducial cross
sections, this approach is not viable due to the complexity of the final state
definition. Instead, one employs a subtraction method to extract infrared
singular real radiation contributions and to recombine them with virtual
contributions to obtain infrared-finite predictions.  The subtracted real and
virtual subprocesses are individually finite and can be integrated numerically,
taking into account the fiducial cuts defining the observable under
consideration. Generic subtraction methods for
NLO~\cite{Catani:1996vz,Frixione:1995ms} and
NNLO~\cite{Binoth:2004jv,Anastasiou:2003gr,Catani:2007vq,Gehrmann-DeRidder:2005btv,Currie:2013vh,Czakon:2010td,Boughezal:2015eha,Gaunt:2015pea,Caola:2017dug,DelDuca:2016ily}
calculations are available and have been used widely for jet cross sections.
For processes involving hadron fragmentation, any subtraction method requires an
extension in order to keep track of parton momentum fractions in unresolved
emissions, which are usually integrated over. Such an extension is available at
NLO for dipole subtraction~\cite{Catani:1996vz}. At NNLO, recent work towards
fragmentation processes yielded results for heavy hadron production in top quark
decays~\cite{Czakon:2021ohs} in the residue subtraction
method~\cite{Czakon:2010td} and photon
fragmentation~\cite{Gehrmann:2022cih,Chen:2022gpk} in the antenna subtraction
method~\cite{Gehrmann-DeRidder:2005btv,Currie:2013vh}.

 It is the objective of this paper to extend the antenna subtraction formalism
 to incorporate hadron fragmentation processes up to NNLO. In
 Section~\ref{sec:notation}, we establish the relevant notation for hadron
 fragmentation processes. Sections~\ref{sec:subNLO} and~\ref{sec:subNNLO}
 develop the antenna subtraction for identified hadrons at NLO and NNLO,
 respectively, by introducing the fragmentation antenna functions and describing
 the structure of the subtraction terms.  The integration of the fragmentation
 antenna functions in final-final kinematics is described in
 Section~\ref{sec:integr}, where we also investigate their relation to inclusive
 antenna functions in initial-final kinematics. Our results are validated by
 re-deriving existing results for single-inclusive NNLO coefficient functions in
 vector boson and Higgs boson decay in Section~\ref{sec:coeffun}.  As an
 illustration of the method, in Section~\ref{sec:epemNLO} we describe the
 subtraction for hadron-in-jet fragmentation in three-jet final states in
 $e^+e^-$ annihilation. Finally, in Section~\ref{sec:concl} we summarise our
 results and provide an outlook on future applications and extensions.

\section{Hadron fragmentation processes in the antenna formalism}
\label{sec:notation}

Processes with identified hadrons require the introduction of a fragmentation
function to describe the fragmentation of the high-energy quark or gluon into
the actually detected hadron.
In this paper, we focus on one hadron (plus jets) production at $\epem$
colliders:
\begin{equation}
  e^+ + e^- \to H(K_H) + X \, (\, + {\rm jets} \,)
\end{equation}
where we identify a hadron $H$ with momentum $K_H$ and possibly some jets, which
may or not contain the identified hadron.
The restriction to $e^+e^-$ initial states is largely for notational simplicity, allowing us the develop the 
essential aspects of the antenna subtraction formalism for identified hadron cross sections in a 
clear and concise manner. Its extension to hadron-hadron collisions is straightforward and will 
be discussed in  Section~\ref{sec:concl}.

The fully differential cross section can be written as
\begin{equation}\label{eq:sigmaH}
  \der\sigma^H = \sum_{p} \int \der\eta\,
  D^H_{p}(\eta,\mu_a^2)\,\der\hsig_{p}(\eta,\mu_a^2)\,,
\end{equation}
where the index $p$ runs over all possible partons in the process, $D^H_p$ is
the physical (mass-factorised) fragmentation function describing the collinear
fragmentation process of the parton $p$ into the hadron $H$, and $\mu_a^2$ is
the fragmentation scale (which may differ from the renormalisation scale).
Note that a single-hadron cross section in QCD is usually written as
differential in the three-momentum of the detected
hadron~\cite{Ellis:1979sj,Aversa:1988vb,Catani:1996vz}  as
\begin{equation}\label{eq:onehadxs}
  K_H^0 \frac{\der \sigma^H}{\der^3 K_H} = \sum_p
  \int \frac{\der\eta}{\eta^2}\, D^H_p(\eta)\, k_p^0
  \left. \frac{\der \hsig_p}{\der^3 k_p} \right|_{\vec{k}_p=\vec{K}_H/\eta}\,,
\end{equation}
where $k_p$ is the momentum of the fragmenting parton, carrying 
$1/\eta$ of the momentum of the detected hadron $K_H$. Such a definition is not
suitable for a parton-level generator. However, as shown in~\cite{Frederix:2018nkq}, \eqref{eq:sigmaH} and~\eqref{eq:onehadxs}
are indeed equivalent for one-particle inclusive cross sections. Hence we will
adopt the former as our master equation, which is necessary to deal with the
additional presence of jets in the final state.

The short-distance one-parton exclusive cross section appearing in~\eqref{eq:sigmaH} admits a
perturbative expansion in the renormalised strong coupling constant $\as$,
\begin{equation}
  \label{eq:sigmaP}  \der\hsig_p(\eta)
  = \der\hsig^{\LO}_p(\eta) + \left( \frac{\as}{2\pi} \right)
  \der\hsig^{\NLO}_p(\eta) + \left( \frac{\as}{2\pi} \right)^2
  \der\hsig^{\NNLO}_p(\eta)\,.
\end{equation}
For instance, the LO cross section is defined as the integration over the $n$
particle phase space of the tree-level Born partonic cross section:
\begin{equation}
  \der\hsig_{p}^{{\rm LO}}(\eta) = \int_{n} \der\hsig_{p}^{\B}(\eta)\,,
\end{equation}
with 
\begin{equation}\label{eq:Bxs}
  \der\hsig_{p}^{\B}(\eta) = {\cal N}_{\B}\,
  \der\Phi_{n}(k_{1},\ldots,k_{n};\q)\,
  \frac{1}{S_{{n}}}\,M^{0}_{n}(k_{1},\ldots,k_{n})\,
  J(\{k_1,\ldots,k_n\}_n, \eta k_p)\,,
\end{equation}
with ${\cal N}_{\B}$ the Born-level normalisation factor, $S_{{n}}$ a symmetry
factor for final-state particles, $M^{0}_{n}$ the squared tree-level
$n$-particle matrix element and $\der\Phi_{n}$ the usual phase space for a
$n$-parton final state with total four-momentum $\q^\mu$ in $d=4-2\e$ space-time
dimensions.
Compared to the standard jet cross sections, the element of novelty here is the
modified jet function $J$, which retains a dependence on the momentum fraction
$\eta$, similarly to what was done in the photon fragmentation case in~\cite{Gehrmann:2022cih}.
The purpose of the modified jet function is to define jet observables and/or any
additional observable depending on the momentum $k_p$ of the identified parton.
In the framework of collinear factorisation encoded in~\eqref{eq:sigmaH},
the momentum $k_p$ of the identified parton is proportional to the momentum
$K_H$ of the identified hadron according to the simple relation $K_H = \eta
k_p$.

Beyond leading order, it is well known that infrared divergences of soft and
collinear origin appear in the short-distance cross section. They are guaranteed
to cancel between real and virtual contributions in sufficiently inclusive
observables, but a subtraction method is required in order to deal with such
divergences in the intermediate steps of the calculation.
In the antenna subtraction
formalism~\cite{Gehrmann-DeRidder:2005btv,Currie:2013vh}, the singularities
associated to single or double unresolved particles in real emission matrix
elements are locally subtracted by means of counterterms built out of antenna
functions~\cite{Gehrmann-DeRidder:2004ttg,Gehrmann-DeRidder:2005svg,Gehrmann-DeRidder:2005alt}. Each
antenna function encodes the radiation pattern between a pair of hard radiators,
thus reproducing the behaviour of the matrix element in the singular limits, but
being simple enough to be analytically integrated over the unresolved degrees of
freedom. The integrated subtraction terms are then added back at the virtual
level, so as to cancel the explicit poles appearing in the virtual matrix
elements.

Most of the elements introduced in~\cite{GehrmannDeRidder:2005cm},
necessary to deal with jet cross sections in $\epem$ collisions, can be 
used for exclusive one-particle cross sections as well.
However, whenever we identify a parton, we spoil the cancellation of collinear
divergences.
The physical reason is that by identifying for example a quark we are in the position
to distinguish a quark from a collinear quark-gluon pair.
These collinear divergences are subtracted from the short-distance cross
sections by means of mass factorisation counterterms and absorbed in the bare
fragmentation functions, which eventually result in mass-factorised
fragmentation functions, the ones appearing in~\eqref{eq:sigmaH}.
In order to allow for a proper subtraction of final-state collinear divergences,
we need to keep track of the momentum fraction of the fragmenting parton in the
intermediate layers of the calculation.
We do so by introducing fragmentation antenna functions which
explicitly depend on the momentum fraction of the fragmenting parton. After integrating over all 
kinematical variables except the momentum fraction, these
fragmentation antenna functions have the proper structure to be combined with
the mass factorisation counterterms and result in a cancellation of final-state
collinear divergence at the integrand level, {\em before} performing the
convolution with the fragmentation function.
In the following sections, we describe how the subtraction has to be modified to
account for the presence of identified hadrons at NLO and NNLO.

\section{Subtraction at NLO}
\label{sec:subNLO}

The NLO corrections to the one-parton exclusive cross section in
\eqref{eq:sigmaP} contain contributions from real emission of one extra parton
and virtual corrections.  As it is customary in antenna subtraction, we
introduce a real subtraction term $\der\hsig_{p}^{\S}$ and a virtual subtraction
term $\der\hsig_{p}^{\T}$, to be subtracted from the real cross section
$\der\hsig_{p}^{\R}$ and the virtual cross section $\der\hsig_{p}^{\V}$,
respectively.
The NLO short-distance cross section can then be written as 
\begin{equation}\label{eq:subNLO}
  \der\hsig_{p}^{\NLO}(\eta) =
  \int_{n+1} \left[ \der\hsig_{p}^{\R}(\eta) - \der\hsig_{p}^{\S}(\eta) \right]
  + \int_{n} \left[ \der\hsig_{p}^{\V}(\eta) - \der\hsig_{p}^{\T}(\eta) \right]\,.
\end{equation}
Each term in square brackets in~\eqref{eq:subNLO} is free of infrared
divergences and suitable for a numerical implementation.
Note the subscript in~\eqref{eq:subNLO}, indicating that each term retains a dependence on
the parton which is undergoing the fragmentation process.

The real partonic cross section $\der\hsig_{p}^{\R}$ is given by
\eqref{eq:Bxs} with an additional parton. It is decomposed according to its colour orderings. 
As for the real subtraction term
$\der\hsig_{p}^{\S}$, it will be given by the sum of several terms, summing over
all possible single unresolved partons:
\begin{equation}\label{eq:sigSNLO}
  \der\hsig_{p}^{\S} = \sum_j \der\hsig_{p,j}^{\S}\,.
\end{equation}
The $\der\hsig_{p,j}^{\S}$ are obtained by summing over all 
colour connections in which the parton $j$ can become unresolved.  
They are further decomposed in two types of contributions as
\begin{equation}
\der\hsig_{p,j}^{\S} = \der\hsig_{p,j}^{\S,{\rm non-id.}p} +  \der\hsig_{p,j}^{\S,{\rm id.}p},  
\end{equation}
where the first term contains all configurations where the identified parton $p$ is not
colour-connected to the unresolved parton $j$, such that  we can use the standard NLO
subtraction term with final-final kinematics, with $p$ appearing unmodified 
in the respective reduced matrix element. 

In order to subtract the infrared limits involving the unresolved parton $j$
colour-connected to the identified $p$ and a second hard parton $k$, we newly
introduce the following subtraction term
\begin{eqnarray}
  \der\hsig_{p,j}^{\S,{\rm id.}p} &=& {\cal N}_{\R}
  \der\Phi_{n+1}(k_{1},\ldots,k_p,\ldots,k_{n+1};\q)\, \frac{1}{S_{{n+1}}}
  \nonumber \\ && \times\,X_3^0(k_p,k_j,k_k)\,
  M^{0}_{n}(k_{1},\ldots,\tilde{K},\tilde{k}_p,\ldots,k_{n+1})\,
  J(\{\ldots,\tilde{K},\tilde{k}_p,\ldots\}_n,\eta\,z\,\tilde{k}_p)\,.
  \label{eq:SNLOxs}
\end{eqnarray}
where ${\cal N}_{\R} = {\cal N}_{\B}\,\overline{C}(\e)/C(\e)$, with
\begin{equation}\label{eq:Ceps}
  C(\e) = \frac{(4\pi e^{-\gamma_E})^\e}{8\pi^2}\,,\quad
  \overline{C}(\e) = (4\pi e^{-\gamma_E})^\e\,,
\end{equation}
which are customary normalisation factors in the antenna subtraction formalism,
and ${\cal N}_{\B}$ the Born-level normalisation factor of the process under
consideration. 
The $X_3^0$ function is just the standard three-particle tree-level antenna
function in the final-final kinematics, depending on the final state momenta
which sum up to $q = k_j+k_k+k_p$ with $\q^2\geq q^2>0$. 
The phase space mapping involves the reconstruction of the momentum fraction
$z$, used to define $\tilde{k}_p = k_p/z$, and of a recoil momentum
$\tilde{K}$. The momentum fraction $z$ is defined by projecting the momentum of
the fragmenting parton and the momentum of its parent parton pair onto a
specific reference four-vector that can be chosen freely.  In our case, we
choose $q$ as reference direction, resulting in
\begin{equation}
  \label{eq:NLOmap}  
  \begin{split}
  z &= \frac{s_{pj}+s_{pk}}{s_{pj}+s_{pk}+s_{jk}}\,, \\
  \tilde{K} &= k_j + k_k - (1-z) \frac{k_p}{z}\,,
  \end{split}
\end{equation}
which satisfies all the required properties. In particular, in the collinear
limit $k_p \parallel k_j$, $z$ approaches the momentum fraction of $k_p$ along
the common collinear direction. Hence the overall momentum fraction entering the
jet function is the product of $\eta$ and $z$. It should be noted that
the definition of $z$ used here differs from the choice made in~\cite{Gehrmann:2022cih}
for photon fragmentation antenna functions, where in the final-final kinematics 
$k_k$ was used as reference in the definition of $z$. As a consequence, the 
integrated NLO fragmentation antenna functions differ from the ones 
listed in~\cite{Gehrmann:2022cih}. The different choice of reference momentum was 
appropriate in the photon case (where emitter and recoil could always be identified in 
an unambiguous manner), but generalises only poorly to the hadron 
fragmentation case. 

In order to reach the factorisation of the phase space, we follow closely
\cite{Daleo:2006xa} by inserting
\begin{equation}\label{eq:ifpm1}
  1=\int \der^d q\,\delta\left(q-k_p-k_j-k_k\right)\,,
\end{equation}
and
\begin{equation}\label{eq:ifpm2}
  1=\frac{q^2}{2\pi}\int\frac{\der z}{z}\int [\der \tilde{K}]\,(2\pi)^{d}
  \delta\left(q - \frac{k_p}{z} - \tilde{K}\right)\,.
\end{equation}
Since in \eqref{eq:SNLOxs} we are integrating over $k_p$, we need to introduce
the one-particle phase space for $\tilde{k}_p$. They are related by
\begin{equation}
  [\der \tilde{k}_p] = [\der k_p]\,z^{2-d} = [\der k_p]\,z^{-2+2\e}\,,
\end{equation}
which is due to the fact that $[\der p] \propto E^{d-3} \der E$. Hence, by
integrating over $q$, we get
\begin{eqnarray}
  \der\Phi_{n+1}(k_{1},\ldots,k_p,k_j,k_k,\ldots,k_{n+1};\q) &=&
  \der\Phi_{n}(k_1,\dots,\tilde{k}_p,\tilde{K},\dots,k_{n+1};\q)\nonumber\\
  &\times&\frac{q^2}{2\pi}\der\Phi_{2}(k_j,k_k;q-k_p)\,z^{1-2\e}\,\der z\,.
\end{eqnarray}
We define the integrated version of the fragmentation antenna function
over the two particle phase space as
\begin{equation}\label{eq:aint}
  \XTZ{X}{p}(z)= \frac{1}{C(\epsilon)}\int \der\Phi_2 \frac{q^2}{2\pi}
  \,z^{1-2\e}\,X_3^0(\id{k}_p,k_j,k_k)\,.
\end{equation}
The (simple) integration is discussed in Section~\ref{sec:X30int}, and explicit
expressions for the $\XTZ{X}{p}$ can be found in Appendix~\ref{app:X30int}.
The integrated form of the subtraction term  is then
\begin{eqnarray}
  \int_1 \der\hsig_{p,j}^{\S,{\rm id.}p} &=& {\cal N}_{\V}\,
  \int \der z\,\der\Phi_{n}(k_1,\dots,\tilde{k}_p,\tilde{K},\dots,k_{n+1};\q)\,
  \frac{1}{S_{n}}
  \nonumber \\ && \times\, {\cal X}_3^{0,{\rm id.} p}(z)
  \,M^{0}_{n}(k_{1},\ldots,\tilde{k}_p,\tilde{K},\ldots,k_{n+1})\,
  J(\{\ldots,\tilde{k}_p,\tilde{K},\ldots\}_n,\eta\,z\,\tilde{k}_p)\,,
  \label{eq:intsigS}
\end{eqnarray}
with ${\cal N}_{\V} = {\cal N}_{\R} C(\e) = {\cal N}_{\B}\,\overline{C}(\e)$.
The above expression contains the infrared poles required to cancel explicit
poles of the virtual matrix element associated with the colour-connections
involving the identified parton momentum $p$ as well as collinear poles
proportional to the Altarelli-Parisi splitting functions, which need to be
properly subtracted by means of a NLO mass factorisation counterterm
$\der\hsig_{p}^{\MF,\NLO}$, defined as
\begin{eqnarray}
  \der\hsig^{\MF,\NLO}_p(\eta) &=& - {\cal N}_{\V}\, \sum_h
  \int \der z\,\der\Phi_{n}(k_1,\dots,k_h,\dots,k_{n};\q)\,
  \mu_a^{-2\e} \MFK{1}{p \rto h}(z)\,
  \nonumber \\ && \times\,  
  \frac{1}{S_{n}}\,M^{0}_{n}(k_{1},\ldots,k_h,\ldots,k_{n})\,
  J(\{k_{1},\ldots,k_h,\ldots,k_{n}\}_n, \eta\,z\,k_h)\,,
\end{eqnarray}
with $\mu_a^2$ the mass factorisation scale and $\MFK{1}{p \rto h}$ the leading
order mass factorisation kernel (see Appendix~\ref{app:mfks}).
The full virtual subtraction term is then given by
\begin{equation}\label{eq:sigT}
  \der\hsig_{p}^{\T}(\eta) = - \sum_j \int_1 \der\hsig_{p,j}^{\S} - \der\hsig_p^{\MF,\NLO}\,.
\end{equation}
Note that the integrated antenna functions and the mass factorisation kernels
can be eventually combined into fragmentation dipoles, analogous to the initial-state dipoles 
introduced for hadron-collider processes~\cite{Currie:2013vh}.

An explicit example of subtraction at NLO is given in Section~\ref{sec:epemNLO},
where we study identified hadron production inside jets in $e^+e^- \to \text{3 jets}$ final states.

\section{Subtraction at NNLO}
\label{sec:subNNLO}

Predictions at NNLO require the calculation of three different contributions,
namely double-real (RR), real-virtual (RV) and double-virtual (VV) contributions
relative to the Born process. Each of these pieces is separately infrared
divergent, whereas their sum is guaranteed to be finite. In order to handle the
implicit divergences and explicit poles that arise in the intermediate steps
of the calculation, three subtraction terms are introduced: a RR subtraction
term $\der\hsig_{p}^{\S}$, a RV subtraction term $\der\hsig_{p}^{\T}$ and a
VV subtraction term $\der\hsig_{p}^{\U}$, such that the analogue of~\eqref{eq:subNLO} at NNLO reads
\begin{equation}\label{eq:subNNLO}
  \der\hsig_{p}^{\NNLO}(\eta) =
  \int_{n+2} \left[ \der\hsig_{p}^{\RR} - \der\hsig_{p}^{\S} \right]
  + \int_{n+1} \left[ \der\hsig_{p}^{\RV} - \der\hsig_{p}^{\T} \right]
  + \int_{n} \left[ \der\hsig_{p}^{\VV} - \der\hsig_{p}^{\U} \right]\,,
\end{equation}
where a dependence on $\eta$ in the integrands is understood.
The structure of the $\der\hsig_{p}^{\S}$, $\der\hsig_{p}^{\T}$ and
$\der\hsig_{p}^{\U}$ subtraction terms has been discussed extensively in
previous works: when the hard radiators are both in the final state (final-final
kinematics)~\cite{GehrmannDeRidder:2005cm}, one radiator in the final state and
one in the initial state (initial-final
kinematics)~\cite{Daleo:2006xa,Daleo:2009yj} or both radiatiors in the initial
state (initial-initial
kinematics)~\cite{Daleo:2006xa,Gehrmann:2011wi,Gehrmann-DeRidder:2012too}. A
comprehensive review of the formalism in hadron-hadron collisions is given in~\cite{Currie:2013vh}.
In this section, we limit ourselves to explain where the subtraction terms have
to been modified in order to account for the presence of the identified
particle.
Since an identified particle in the final state is conceptually similar to an
initial state particle, the structure of the subtraction terms is close to the
one provided in the initial-final case in~\cite{Daleo:2009yj}.

The real-real subtraction term $\der\hsig_{p}^{\S}$ is built out of several
pieces, each of which accounts for a particular type of unresolved
configuration.
The first piece, $\der\hsig_{p}^{\S, a}$ deals with the the single unresolved
limits of the double real matrix elements. Its structure is similar to the NLO
real subtration term, already introduced in~\eqref{eq:sigSNLO}.
The second piece, $\der\hsig_{p}^{\S, b}$ accounts for the double unresolved
limits of the RR matrix element. We can distinguish configurations with the
identified parton $p$ colour-connected or not to the pair of unresolved partons
$j$ and $k$. In the latter case, we can use the standard NNLO subtraction
term. In the former case, we introduce the subtraction term 
\begin{eqnarray}
\lefteqn{  \der\hsig_{p,jk}^{\S, b,{\rm id.}p} = {\cal N}_{\RR}\,
  \der\Phi_{n+2}(k_{1},\ldots,k_{n+2};\q)\, \frac{1}{S_{{n+2}}}  }
  \nonumber \\ &&\times
  \bigg[X_4^0(k_p,k_j,k_k,k_l)   M^{0}_{n}(k_{1},\ldots,{\tilde{k}}_p,{\tilde{K}},\ldots,k_{n+2})\,
  J(\{\ldots,{\tilde{k}}_p,{\tilde{K}},\ldots\}_n,\eta\,z\,{\tilde{k}}_p)
  \nonumber \\ &&
      - X_3^0(k_p,k_j,k_k)\,X_3^0(\tilde{k}_p,\tilde{K},k_l) M^{0}_{n}(k_{1},\ldots,\tilde{\tilde{k}}_p,\tilde{\tilde{K}},\ldots,k_{n+2})\,
  J(\{\ldots,\tilde{\tilde{k}}_p,\tilde{\tilde{K}},\ldots\}_n,\eta\,z\,\tilde{\tilde{k}}_p)   \nonumber \\ &&
    - X_3^0(k_j,k_k,k_l)\,X_3^0(k_p,\tilde{k}_{jk},\tilde{k}_{kl}) 
    M^{0}_{n}(k_{1},\ldots,\tilde{\tilde{k}}_p,\tilde{\tilde{K}},\ldots,k_{n+2})\,
  J(\{\ldots,\tilde{\tilde{k}}_p,\tilde{\tilde{K}},\ldots\}_n,\eta\,z\,\tilde{\tilde{k}}_p)
    \bigg]
\,,\nonumber \\
  \label{eq:sigSb}
\end{eqnarray}
with ${\cal N}_{\RR} = {\cal N}_{\B}\,\overline{C}(\e)^2/C(\e)^2$.
The two products of three-parton antenna function are necessary to subtract the
single unresolved limits of the four-parton antenna function, such that
$\der\hsig_{p,jk}^{\S, b,{\rm id.}p}$ is active only in the double unresolved
limits.  They each involve two consecutive NLO phase space mappings, whose
results are abbreviated as $(\tilde{\tilde{k}}_p,\tilde{\tilde{K}})$ and where
the intermediate momenta $(\tilde{k}_{jk},\tilde{k}_{kl})$ indicate a standard
final-final NLO mapping.
The genuine NNLO mapping to $({\tilde{k}}_p = k_p/z,{\tilde{K}})$ in the first
term is a generalisation of~\eqref{eq:NLOmap} with more than one parton becoming
unresolved. Explicitly, it reads:
\begin{equation}
  \label{eq:NNLOmap}  
  \begin{split}
    z &= \frac{s_{pj}+s_{pk}+s_{pl}}{s_{pj}+s_{pk}+s_{jk}+s_{pl}+s_{jl}+s_{kl}}\,,
    \\ {\tilde{K}} &= k_j + k_k + k_l - (1-z) \frac{k_p}{z}\,.    
  \end{split}
\end{equation}
with $p$ the fragmenting parton, $j$ and $k$ the two unresolved partons, and $l$
the other final state radiator. Such a mapping satisfies the appropriate limits in
all double singular configurations. 
Moreover, it turns into an NLO phase space mapping in its single unresolved
limits, as required in order to cancel the single unresolved limits of the
$X_4^0$ antenna function.
The integral of the tree-level four-particle antenna function over the
three-particle phase space
\begin{equation}\label{eq:X40int}
  \XFZ{X}{p}(z) = \frac{1}{[C(\e)]^2}\int \der\Phi_3 \frac{q^2}{2\pi}
  \,z^{1-2\e}\,X_4^0(\id{k}_p,k_j,k_k,k_l)\,,
\end{equation}
reappears at the virtual-virtual level; its integration is discussed in
Section~\ref{sec:X40int}.

The subtraction terms with two unresolved partons almost colour-unconnected
($\der\hsig^{\S, c}$) or colour-unconnected ($\der\hsig^{\S, d}$) do not require
new ingredients: they contain products of tree-level three-parton antenna
functions or fragmentation antenna functions in the final-final
kinematics.
They appear, after integration, at the real-virtual or at the virtual-virtual
level, as the product of an integrated antenna function $\mathcal{X}_3^0$ with an
unintegrated antenna function $X_3^0$ or as a product of two $\mathcal{X}_3^0$,
respectively.
In order to avoid oversubtraction of large-angle soft gluon radiation, 
additional soft antenna functions $S_{ajc}$~\cite{Gehrmann-DeRidder:2007foh} are
used to construct $\der\hsig^{\S, e}$; in such soft antenna functions, the hard momenta $a$
and $c$ can be arbitrary on-shell momenta in the initial or final state.
The integral of the soft antenna function in the final-final kinematics is given
in~\cite{Gehrmann-DeRidder:2007foh}, whereas in the initial-final
kinematics in~\cite{Daleo:2009yj,Currie:2013vh}.
Given the freedom we have in the choice of the hard momenta of the soft antenna
function, we can use the known results in processes with fragmentation as well.

At the real-virtual level, we need to remove the explicit infrared poles of the
one-loop matrix element and to also subtract its single unresolved limit.
The former purpose is accomplished by the integral of $\der\hsig_{p}^{\S, a}$,
which combined with mass factorisation terms (see below), results in
$\der\hsig_{p}^{\T, a}$.
The latter purpose requires the introduction of a new subtraction term:
\begin{eqnarray}
  \der\hsig_{p,j}^{\T, b,{\rm id.}p} &=& {\cal N}_{\RV}\,
  \der\Phi_{n+1}(k_{1},\ldots,k_{n+1};\q)\, \frac{1}{S_{{n+1}}}
  J(\{\ldots,\tilde{k}_p,\tilde{K},\ldots\}_n,\eta\,z\,\tilde{k}_p)  
  \nonumber \\ &&
  \times\,\Bigg[ X_3^0(k_p,k_j,k_k)\,
    M^{1}_{n}(k_{1},\ldots,\tilde{k}_p,\tilde{K},\ldots,k_{n+1})\,
  \nonumber \\ && \phantom{\times\Bigg[}
    + X_3^1(k_p,k_j,k_k)\,
    M^{0}_{n}(k_{1},\ldots,\tilde{k}_p,\tilde{K},\ldots,k_{n+1})
    \Bigg]\,.
  \label{eq:sigTb}
\end{eqnarray}
with ${\cal N}_{\RV} = {\cal N}_{\B}\,\overline{C}(\e)^2/C(\e)$, and where we
have used the NLO momentum mapping~\eqref{eq:NLOmap}.
In here, $M^{1}_{n}$ is the one-loop reduced matrix element and $X_3^1$ is the
one-loop three-parton antenna function in the final-final kinematics, whose
integral over the two-particle phase space is denoted as
\begin{equation}
  \XTO{X}{p}(z) = \frac{1}{C(\e)}\int \der\Phi_2 \frac{q^2}{2\pi}
  \,z^{1-2\e}\,X_3^1(\id{k}_p,k_j,k_k)\,;
\end{equation}
it is reintroduced at the virtual-virtual level. Its integration is presented in
Section~\ref{sec:X31int}.
In order to assemble $\der\hsig_{p}^{\T, a}$ and $\der\hsig_{p}^{\T, b}$, one also
needs the real-virtual mass factorisation term:
\begin{equation}
  \der\hsig^{\MF,\RV}_p(\eta) = - {\cal N}_{\RV}\,\sum_h  
  \int \der z\, \mu_a^{-2\e} \MFK{1}{p \rto h}(z)
   \left( \der\hsig^{\R}_h(\eta\,z) - \der\hsig^{\S}_h(\eta\,z) \right) \,.
\end{equation}
which contributes both to $\der\hsig_{p}^{\T, a}$ and $\der\hsig_{p}^{\T, b}$.
The last piece needed for the subtraction at the real-virtual level is
$\der\hsig^{\T, c}$, which results from the integration of $\der\hsig^{\S, c}$
and $\der\hsig^{\S, e}$, plus additional terms to ensure an IR finite
contribution, which are added back at the double virtual level.

At the virtual-virtual level, there are no implicit infrared divergences; the
explicit poles of the two-loop matrix element are canceled by the integrated
form of the appropiate subtraction terms, together with the double virtual mass
factorisation term:
\begin{eqnarray}
  \der\hsig^{\MF,\VV}_p(\eta) &=& - {\cal N}_{\VV}\,\sum_h  
  \int \der z\,\mu_a^{-2\e} \Bigg[ \MFK{2}{p \rto h}(z)\,
    \der\hsig^{\B}_h(\eta\,z)
    \nonumber \\ && \phantom{- {\cal N}_{\VV}\, \int \der z\, \sum_h \Bigg[ \mu_a^{-2\e}}
    + \MFK{1}{p \rto h}(z)\,
    \Big( \der\hsig^{\V}_h(\eta\,z) - \der\hsig^{\T}_h(\eta\,z) \Big)
    \Bigg]  \,.
    \label{eq:MFVV}
\end{eqnarray}
with ${\cal N}_{\VV} = {\cal N}_{\B}\,\overline{C}(\e)^2$, to result in
$\der\hsig_{p}^{\U}$, and $\MFK{2}{p \rto h}$ are the colour-stripped version of
the next-to-leading order splitting kernels defined in~\eqref{eq:MF2}.

\section{Integration of fragmentation antenna functions}
\label{sec:integr}

The integration of the $X_3^0$, $X_4^0$ and $X_3^1$ fragmentation antenna
functions is closely related to the integration of the corresponding
initial-final antenna functions described in~\cite{Daleo:2006xa,Daleo:2009yj}.

\subsection{Integration of NLO antenna functions}
\label{sec:X30int}

According to the definition of the ${\cal X}_3^{0,{\rm id.} p}$ given in
\eqref{eq:aint}, we integrate the antenna function over the two-particle
phase space with kinematics
\begin{equation}\label{eq:2to2kin0}
  q + (-k_p) \to k_1 + k_2\,,
\end{equation}
with $s_{12} = (q-k_p)^2 = q^2 (1-z)$ and
\begin{equation}\label{eq:zdef}
  z = \frac{2\,k_p \cdot q}{q^2}\,.
\end{equation}
The fragmenting parton $k_p$ may be
regarded as an initial state parton with negative four-momentum, and we are
then looking at the $1 \to 3$ process with one identified parton as a $2 \to
2$ scattering process with rescaled invariant mass.
By inserting the explicit expression for the two-particle phase space,
\eqref{eq:2to2kin0} reduces to a simple one-dimensional integral
\begin{equation}
  {\cal X}_3^{0,{\rm id.} p}(z) = z^{1-2\e} (1-z)^{-\e} \left( \frac{q^2}{4} \right)^{1-\e}
  \frac{e^{\gamma \e}}{\Gamma (1-\e)} \int_{-1}^{+1}\der v\, (1-v^2)^{-\e}\,
  X_3^0 (s_{12},s_{1p},s_{2p}) 
\end{equation}
where 
\begin{equation}
  s_{1p} = \frac{q^2}{2} z (1-v)\,,\quad s_{2p} = \frac{q^2}{2} z (1+v)\,.
\end{equation}
After integration, the ${{(1-z)}^{-\e}}$ factor, which
regulates end-point soft divergences, can be safely expanded in term of
distributions, according to
\begin{equation}\label{eq:expdistr}
  (1-z)^{-1+k\e} = -\frac{1}{k\e}\delta(1-z)
  + \sum_{n} \frac{(-k\e)^n}{n!} \Dn{n}(1-z)
\end{equation}
with
\begin{equation}
  \Dn{n}(1-z) = \left( \frac{\log^n(1-z)}{1-z} \right)_+\,.
\end{equation}
Explicit expressions for the integrated ${\cal X}_3^{0,{\rm id.} p}$ antenna
functions are provided in Appendix~\ref{app:X30int}.  We note that they can be
related to the inclusively integrated initial-final antenna functions derived
in~\cite{Daleo:2006xa} by replacing
\begin{equation}
\label{eq:STcont}  Q^2 \to - q^2\,,\quad x \to 1/z\,.
\end{equation}
as can be evidenced form \eqref{eq:2to2kin0}. 

\subsection{Integration of NNLO real-real antenna functions}
\label{sec:X40int}

Similarly to the steps preformed in the previous section, we integrate the
$X_4^0$ over a three-particle phase space with $2 \to 3$ kinematics
\begin{equation}\label{eq:2to3kin0}
  q + (-k_p) \to k_1 + k_2 + k_3\,.
\end{equation}
The integration is performed with well-known techniques, based on multi-loop
calculation technology. Namely, we rewrite the three-particle phase space in
terms of cut propagators, in order to express it as a cut through a three-loop
vacuum polarisation diagram. Then, we reduce the occurring integrals as linear
combinations of a smaller set of master integrals, with the help of
\textsc{Reduze2}~\cite{vonManteuffel:2012np}. We managed to reduce to the same
set of 9 master integrals which appears in the initial-final case, see Section~4 of~\cite{Daleo:2009yj}.
This is ultimately due to the fact that the processes
\begin{equation}\label{eq:kinTL}
  q + (-k_p) \to k_1 + k_2 + k_3\,,\quad q^2 > 0\,,
  \quad (q-k_p)^2 = (1-z)\,q^2
\end{equation}
and
\begin{equation}\label{eq:kinSL}
  q + k_i \to k_1 + k_2 + k_3\,,\quad q^2 = - Q^2  < 0\,,
  \quad (q+k_i)^2 = (1-x)/x\,Q^2
\end{equation}
feature the same kinematics with a different definition of invariants.

The master integrals have been calculated by exploiting the differential
equations method~\cite{Gehrmann:1999as}. We first use
\textsc{epsilon}~\cite{Prausa:2017ltv} to find the canonical
form~\cite{Henn:2014qga,Henn:2013pwa} of the differential equations. Once
expressed in the canonical form, the differential equations can be iteratively
solved in term of the harmonic polylogarithms (HPLs)~\cite{Remiddi:1999ew}, with
unknown boundary conditions.
In order to impose the boundary conditions, and at the same time check the
structure of the master integrals, we can fully exploit the similarity between
\eqref{eq:kinTL} and \eqref{eq:kinSL}: the master integrals for
real-real fragmentation antenna functions are related to the master integrals
for initial-final antenna functions, reported in Appendix~A.1 of~\cite{Daleo:2009yj}, 
by means of the replacement \eqref{eq:STcont}, 
which amounts to an analytic continuation. In particular, HPLs of
argument $1/z$ are expressed as HPLs of argument $z$ by means of an iterative
procedure (see for instance Section~6 of~\cite{Remiddi:1999ew}), which is
implemented in the Mathematica package
\textsc{HPL}~\cite{Maitre:2005uu,Maitre:2007kp}. Full consistency has been found
with the master integrals found by means of the differential equation
approach, after adjustment of some typographical mistakes, in particular  
 in (A.9) of~\cite{Daleo:2009yj} the endpoint term should be corrected as
\begin{displaymath}
  \e \left( 256\,\zeta_3 - \frac{43}{18}\pi^4 \right)
  \to \e \left( -\frac{74}{45}\pi^4 \right)\,.
\end{displaymath}
Once the explicit expressions for the master integrals have been found, they can
be substituted inside the antenna functions. At this point, the factor
$(1-z)^{-2\e}$ can be safely expanded in distributions according to
\eqref{eq:expdistr}. The whole workflow of the calculation has been implemented in
\textsc{FORM}~\cite{Vermaseren:2000nd}.

In Table~\ref{tab:X40} we list all the integrated tree-level four-parton antenna
functions; they differ by the nature of the identified particle and the hard
radiators they collapse to.
Some momentum permutations of identified particles in $\XFZ{A}{p}$,
$\XFZ{B}{p}$, $\XFZ{D}{p}$, $\XFZ{\tilde{E}}{p}$ and in all gluon-gluon antenna
functions are not shown in Table~\ref{tab:X40}, because the integrated antenna
functions turn out to be the same: this is ultimately due to the symmetries
present in the unintegrated antenna functions under permutation of partons of
the same flavour.
Instead in the case of $\XFZ{C}{p}$ and $\XFZ{E}{p}$ different identified
particles lead to different results at the integrated level: hence they are
distinguished by a label indicating the fragmenting parton.
Explicit expressions for the integrated $\XFZ{X}{p}$ antenna functions are
provided as ancillary files to the arXiv submission of the paper.

\begin{table}
\centering
\begin{tabular}{c c c c c}
  \toprule
  Hard radiators & Notation & Integral of & In the ancillary file \\
  \midrule
  \multirow{10}{*}{Quark-quark}
  & $\XFZ{A}{q}$ & $A_4^0(\id{1}_q, 3_g, 4_g, 2_{\qb})$ & \texttt{qA40} \\ 
  & $\XFZ{A}{g}$ & $A_4^0(1_q, \id{3}_g, 4_g, 2_{\qb})$ & \texttt{gA40} \\
  \cmidrule{2-4}
  & $\XFZ{\tilde{A}}{q}$ & $\tilde{A}_4^0(\id{1}_q, 3_g, 4_g, 2_{\qb})$ & \texttt{qA40t} \\ 
  & $\XFZ{\tilde{A}}{g}$ & $\tilde{A}_4^0(1_q, \id{3}_g, 4_g, 2_{\qb})$ & \texttt{gA40t} \\ 
  \cmidrule{2-4}
  & $\XFZ{B}{q}$   & $B_4^0(\id{1}_q, 3_{\qp}, 4_{\qbp}, 2_{\qb})$ & \texttt{qB40} \\ 
  & $\XFZ{B}{\qp}$ & $B_4^0(1_q, \id{3}_{\qp}, 4_{\qbp}, 2_{\qb})$ & \texttt{qpB40} \\  
  \cmidrule{2-4}
  & $\XFZ{C}{\qb}$ & $C_4^0(1_q, 3_{q}, 4_{\qb}, \id{2}_{\qb})$ & \texttt{qbC40} \\ 
  & $\XFZ{C}{q1}$  & $C_4^0(\id{1}_q, 3_{q}, 4_{\qb}, 2_{\qb})$ & \texttt{q1C40} \\
  & $\XFZ{C}{q3}$  & $C_4^0(1_q, \id{3}_{q}, 4_{\qb}, 2_{\qb})$ & \texttt{q3C40} \\    
  \midrule  
  \multirow{11}{*}{Quark-gluon}
  & $\XFZ{D}{q}$  & $D_4^0(\id{1}_q, 2_g, 3_g, 4_g)$ & \texttt{qD40} \\ 
  & $\XFZ{D}{g2}$ & $D_4^0(1_q, \id{2}_g, 3_g, 4_g)$ & \texttt{g2D40} \\
  & $\XFZ{D}{g3}$ & $D_4^0(1_q, 2_g, \id{3}_g, 4_g)$ & \texttt{g3D40} \\
  \cmidrule{2-4}
  & $\XFZ{E}{g}$ & $E_4^0(1_q, 2_{\qp}, 3_{\qbp}, \id{4}_g)$ & \texttt{gE40} \\
  & $\XFZ{E}{q1}$ & $E_4^0(\id{1}_q, 2_{\qp}, 3_{\qbp}, 4_g)$ & \texttt{q1E40} \\
  & $\XFZ{E}{q2}$ & $E_4^0(1_q, \id{2}_{\qp}, 3_{\qbp}, 4_g)$ & \texttt{q2E40} \\
  & $\XFZ{E}{q3}$ & $E_4^0(1_q, 2_{\qp}, \id{3}_{\qbp}, 4_g)$ & \texttt{q3E40} \\ 
  \cmidrule{2-4}
  & $\XFZ{\tilde{E}}{g}$ & $\tilde{E}_4^0(1_q, 2_{\qp}, 3_{\qbp}, \id{4}_g)$ & \texttt{gE40t} \\
  & $\XFZ{\tilde{E}}{q1}$ & $\tilde{E}_4^0(\id{1}_q, 2_{\qp}, 3_{\qbp}, 4_g)$ & \texttt{q1E40t} \\
  & $\XFZ{\tilde{E}}{q2}$ & $\tilde{E}_4^0(1_q, \id{2}_{\qp}, 3_{\qbp}, 4_g)$ & \texttt{q2E40t} \\
  \midrule
  \multirow{7}{*}{Gluon-gluon}
  & $\XFZ{F}{g}$ & $F_4^0(1_g, \id{3}_g, 4_g, 2_{g})$ & \texttt{gF40} \\
  \cmidrule{2-4}  
  & $\XFZ{G}{g}$ & $G_4^0(\id{1}_g, 3_{q}, 4_{\qp}, 2_g)$ & \texttt{gG40} \\ 
  & $\XFZ{G}{q}$ & $G_4^0(1_g, \id{3}_{q}, 4_{\qp}, 2_g)$ & \texttt{qG40} \\
  \cmidrule{2-4}
  & $\XFZ{\tilde{G}}{g}$ & $\tilde{G}_4^0(\id{1}_g, 3_{q}, 4_{\qp}, 2_g)$ & \texttt{gG40t} \\ 
  & $\XFZ{\tilde{G}}{q}$ & $\tilde{G}_4^0(1_g, \id{3}_{q}, 4_{\qp}, 2_g)$ & \texttt{qG40t} \\
  \cmidrule{2-4}  
  & $\XFZ{H}{q}$ & $\hat{H}_4^0(\id{1}_q, 3_{\qb}, 4_{\qp}, 2_{\qbp})$ & \texttt{qH40} \\
  \bottomrule
\end{tabular}
\caption{Integrated tree-level four-parton antenna functions $\XFZ{X}{p}$.}
\label{tab:X40}
\end{table}

\subsection{Integration of NNLO real-virtual antenna functions}
\label{sec:X31int}

The integration of the $X_3^1$ fragmentation antenna is performed over a
two-particle phase space with $2 \to 2$ kinematics, as in Section~\ref{sec:X30int}.
The $X_3^1$ antenna functions are expressed in terms of rational functions of
invariants multiplying one-loop bubble and box integrals.
In order to use the same techniques of Section~\ref{sec:X40int}, we rewrite the
one-loop integrals in terms of propagators, and then we write the two-particle
phase space integral of the one-loop antenna functions as a three-loop integral
with two cut propagators.
By doing so, we reduce to the same set of 6 master integrals of the
initial-final case~\cite{Daleo:2009yj}. The master integrals are determined with the differential
equation method, with boundary conditions obtained by internal consistency of
the set of equations, or by a direct calculation at $z=1$.
Three of them contain as subdiagram a one-loop bubble, one of them a
one-loop triangle (that can be expressed in terms of one-loop bubbles), and two
of them a one-loop box.

In the real-virtual case, the master integrals for fragmentation antenna
functions cannot be inferred from the master integrals for initial-final antenna
functions, reported in Appendix~A.2 of~\cite{Daleo:2009yj}, since the
analytic continuation from \eqref{eq:kinSL} to \eqref{eq:kinTL} acts
differently on the different bubble and box integrals and must be performed
prior to the phase space integration. Consequently, a simple relationship
between space-like and time-like real-virtual master integrals can not be
established.

The integrated one-loop squared matrix elements are subsequently renormalised, 
as described in detail in Section 4.2 of~\cite{Daleo:2009yj}:
the strong coupling constant renormalisation is carried out in the
$\overline{{\rm MS}}$ scheme at fixed scale $\mu^2 = q^2$; in the case of
quark-gluon and gluon-gluon antenna functions, the effective operators used to
couple an external current to the parton radiators are also renormalised.
Finally, in order to obtain the integrated one-loop antenna functions, we subtract
from the renormalised one-loop squared matrix elements the corresponding
integrated tree-level antenna function multiplied with the virtual one-loop
correction to the hard radiator vertex.

In Table~\ref{tab:X31} we list all the integrated one-loop three-parton antenna
functions; they differ by the nature of the identified particle and the hard
radiators they collapse to.
Explicit expressions for the integrated $\XTO{X}{p}$ antenna functions are
provided as ancillary files to the arXiv submission of the paper.

\begin{table}
\centering
\begin{tabular}{c c c c}
  \toprule
  Hard radiators & Notation & Integral of & In the ancillary file \\
  \midrule
  \multirow{6}{*}{Quark-quark}
  & $\XTO{A}{q}$ & $A_3^1(\id{1}_q, 3_g, 2_{\qb})$ & \texttt{qA31} \\ 
  & $\XTO{A}{g}$ & $A_3^1(1_q, \id{3}_g, 2_{\qb})$ & \texttt{gA31} \\
  \cmidrule{2-4}
  & $\XTO{\tilde{A}}{q}$ & $\tilde{A}_3^1(\id{1}_q, 3_g, 2_{\qb})$ & \texttt{qA31t} \\ 
  & $\XTO{\tilde{A}}{g}$ & $\tilde{A}_3^1(1_q, \id{3}_g, 2_{\qb})$ & \texttt{gA31t} \\ 
  \cmidrule{2-4}
  & $\XTO{\hat{A}}{q}$ & $\hat{A}_3^1(\id{1}_q, 3_g, 2_{\qb})$ & \texttt{qA31h} \\ 
  & $\XTO{\hat{A}}{g}$ &  $\hat{A}_3^1(1_q, \id{3}_g, 2_{\qb})$ & \texttt{gA31h} \\
  \midrule
  \multirow{12}{*}{Quark-gluon}
  & $\XTO{D}{q}$ & $D_3^1(\id{1}_q, 3_g, 2_{g})$ & \texttt{qD31} \\ 
  & $\XTO{D}{g}$ & $D_3^1(1_q, \id{3}_g, 2_{g})$ & \texttt{gD31} \\
  \cmidrule{2-4}
  & $\XTO{\hat{D}}{q}$ & $\hat{D}_3^1(\id{1}_q, 3_g, 2_{g})$ & \texttt{qD31h} \\ 
  & $\XTO{\hat{D}}{g}$ & $\hat{D}_3^1(1_q, \id{3}_g, 2_{g})$ & \texttt{gD31h} \\
  \cmidrule{2-4}  
  & $\XTO{E}{q}$ & $E_3^1(\id{1}_q, 3_{\qp}, 2_{\qbp})$ & \texttt{qE31} \\ 
  & $\XTO{E}{\qp}$ & $E_3^1(1_q, \id{3}_{\qp}, 2_{\qbp})$ & \texttt{qpE31} \\
  \cmidrule{2-4}
  & $\XTO{\tilde{E}}{q}$ & $\tilde{E}_3^1(\id{1}_q, 3_{\qp}, 2_{\qbp})$ & \texttt{qE31t} \\ 
  & $\XTO{\tilde{E}}{\qp}$ & $\tilde{E}_3^1(1_q, \id{3}_{\qp}, 2_{\qbp})$ & \texttt{qpE31t} \\
  \cmidrule{2-4}  
  & $\XTO{\hat{E}}{q}$ & $\hat{E}_3^1(\id{1}_q, 3_{\qp}, 2_{\qbp})$ & \texttt{qE31h} \\ 
  & $\XTO{\hat{E}}{\qp}$ & $\hat{E}_3^1(1_q, \id{3}_{\qp}, 2_{\qbp})$ & \texttt{qpE31h} \\
  \midrule
  \multirow{10}{*}{Gluon-gluon}
  & $\XTO{F}{g}$ & $F_3^1(1_g, \id{3}_g, 2_{g})$ & \texttt{gF31} \\
  \cmidrule{2-4}
  & $\XTO{\hat{F}}{g}$ & $\hat{F}_3^1(1_g, \id{3}_g, 2_{g})$ & \texttt{gF31h} \\
  \cmidrule{2-4}  
  & $\XTO{G}{g}$ & $G_3^1(\id{1}_g, 3_{q}, 2_{\qp})$ & \texttt{gG31} \\ 
  & $\XTO{G}{q}$ & $G_3^1(1_g, \id{3}_{q}, 2_{\qp})$ & \texttt{qG31} \\
  \cmidrule{2-4}
  & $\XTO{\tilde{G}}{g}$ & $\tilde{G}_3^1(\id{1}_g, 3_{q}, 2_{\qp})$ & \texttt{gG31t} \\ 
  & $\XTO{\tilde{G}}{q}$ & $\tilde{G}_3^1(1_g, \id{3}_{q}, 2_{\qp})$ & \texttt{qG31t} \\
  \cmidrule{2-4}  
  & $\XTO{\hat{G}}{g}$ & $\hat{G}_3^1(\id{1}_g, 3_{q}, 2_{\qp})$ & \texttt{gG31h} \\ 
  & $\XTO{\hat{G}}{q}$ & $\hat{G}_3^1(1_g, \id{3}_{q}, 2_{\qp})$ & \texttt{qG31h} \\
  \bottomrule
\end{tabular}
\caption{Integrated one-loop three-parton antenna functions $\XTO{X}{p}$.}
\label{tab:X31}
\end{table}

\section{Coefficient functions for hadron production at $\epem$ colliders}
\label{sec:coeffun}

\subsection{Identified hadrons in $\gamma/Z$ boson decay}
\label{sec:cfgamma}

Next-to-next-to-leading order corrections to the coefficient functions
contributing to the longitudinal and transverse one-hadron energy spectrum in
$\epem$ annihilation have been first derived in~\cite{Rijken:1996vr,Rijken:1996ns} and independently rederived in~\cite{Mitov:2006ic}. Since the antenna functions are extracted from
double-real and real-virtual matrix elements for $\gamma^* \to q\bar{q}$
decay~\cite{Gehrmann-DeRidder:2004ttg}, by combining integrated antenna
functions with quark form factors, we are in the position to calculate these 
coefficient functions in an independent manner. This comparison provides a
strong check on the correctness of our integrated $\mathcal{A}$-type,
$\mathcal{B}$-type and $\mathcal{C}$-type fragmentation antenna functions, and
indirectly on our whole procedure.

The cross section differential in the energy fraction $x = 2 E_p/\sqrt{s}$ of
the identified hadron is usually written as
\begin{equation}
  \frac{\der\sigma^H}{\der x}
  = \int_{x}^1 \frac{\der z}{z}\,
  \sum_{p=1}^{\NF} \sigma^{(0)}_{p}
  \left[
  D_{\S}^H\left( \frac{x}{z} \right) \mathbb{C}^{\S}_q(z)
  + D_g^H\left( \frac{x}{z} \right) \mathbb{C}^{\S}_g(z) 
  + D_{\NS,p}^H\left( \frac{x}{z} \right) \mathbb{C}^{\NS}_{q}(z)
  \right]\,,
\end{equation}
where we have introduced the singlet (S) and non-singlet (NS) combination of
fragmentation densities, defined as
\begin{equation}
  D_{\S}^H = \frac{1}{N_F} \sum_{p=1}^{N_F} \left( D^H_p + D^H_{\bar{p}} \right)\,,
  \quad
  D_{\NS,p}^H = D^H_p + D^H_{\bar{p}} - D_{\S}^H\,.
\end{equation}
It is customary to also define the purely singlet (PS) coefficient function
$\mathbb{C}^{\PS}_{q} = \mathbb{C}^{\S}_q - \mathbb{C}^{\NS}_{q}$.

Higher order QCD corrections to the $\mathbb{C}$ coefficient functions originate
from radiation in the final state.
All electroweak effects factor out of the coefficient functions, and are
included in the pointlike total cross section $\sigma^{(0)}_{p}$ for the process
$e^+ + e^- \to p + \bar{p}$, which in the simple QED-only case is equal to the
well-known $e_p^2 N 4 \pi \alpha^2/(3 s)$.

The coefficient functions $\mathbb{C}$ are the mass-factorised and
UV-renormalised version of the parton fragmentation functions $\ff$, whose
expansion in the (unrenormalised) strong coupling constant reads
\begin{equation}\label{eq:ffexp}
  \hat{\mathcal{F}} = \ff^{(0)}
  + \left(\frac{\hat{\alpha}_s}{4\pi} \right) S_\e\,
  \left( \frac{\mu_0^2}{Q^2} \right)^\e  \ff^{(1)}
  + \left(\frac{\hat{\alpha}_s}{4\pi} \right)^2 S^2_\e\,
  \left( \frac{\mu_0^2}{Q^2} \right)^{2\e}  \ff^{(2)}
  + \mathcal{O}(\hat{\alpha}_s^3)\,,
\end{equation}
where $\hat{\alpha}_s$ is the bare coupling, $S_\e = (4\pi e^{-\gamma_E})^\e$
and $\mu_0^2$ is the mass parameter of dimensional regularisation.
The parton fragmentation functions are obtained as projections of the parton
structure tensor $\hat{W}_{\mu\nu}$ onto its longitudinal ($\ff_L$), transverse
($\ff_T$) and asymmetric ($\ff_A$) components. They are in
one-to-one correspondence with the homonymous contributions in the angular
distribution of the detected hadron,
\begin{equation}
  \frac{\der^2\sigma^H}{\der x\, \der \cos\theta}
  = \frac{3}{8}(1+\cos^2\theta) \frac{\der\sigma^H_T}{\der x}
  + \frac{3}{4}\sin^2\theta \frac{\der\sigma^H_L}{\der x}
  + \frac{3}{4}\cos\theta \frac{\der\sigma^H_A}{\der x}\,.
\end{equation}
Since we are fully inclusive over the angle of emission of the detected hadron,
we consider the trace $\hat{W}^\mu_\mu$, which is related to the combination
\begin{equation}
  -\frac{z}{d-2} \hat{W}^\mu_\mu = \ff_T + \frac{2}{d-2} \ff_L \equiv \ff_U\,.
\end{equation}
The chosen normalisation is such that, at leading order, we have
\begin{equation}
  \ff^{(0)}_{U,q} = \ff^{(0)}_{T,q} = \delta(1-z)\,,\quad \ff^{(0)}_{L,q} = 0\,,\quad
  \ff^{(0)}_{U,g} = \ff^{(0)}_{T,g} = \ff^{(0)}_{L,g} = 0\,.
\end{equation}
The relationships we observe between the parton fragmentation functions and the
integrated fragmentation antenna functions read at NLO
\begin{align}
  \frac{1}{C_F} \ff_{U,q}^{(1)} &= 4\,\mathcal{A}_{3}^{0,{\rm id.}q}
  + 8\,\delta(1-z) V^{(1)}_q|_{N}\,, \\
  \frac{1}{C_F} \ff_{U,g}^{(1)} &= 4\,\mathcal{A}_{3}^{0,{\rm id.}g}\,,
\end{align}
and at NNLO
\begin{eqnarray}
  \frac{1}{C_F} \ff_{U,q}^{\NS,(2)}|_{N} &=& 2\,\mathcal{A}_{4}^{0,{\rm id.}q}
  + 8\,\mathcal{A}_{3}^{1U,{\rm id.} q} + \delta(1-z) V^{(2)}_q|_{N}\,, \\
  \frac{1}{C_F} \ff_{U,q}^{\NS,(2)}|_{N_F} &=& 2\,\mathcal{B}_{4}^{0,{\rm id.}q}
  + \delta(1-z) V^{(2)}_q|_{N_F}\,, \\
  \frac{1}{C_F} \ff_{U,q}^{\NS,(2)}|_{1/N} &=& -\tilde{\mathcal{A}}_{4}^{0,{\rm id.}q}
  - 8\,\tilde{\mathcal{A}}_{3}^{1U,{\rm id.} q}
  - 4\,\mathcal{C}_{4}^{0,{\rm id.}\bar{q}}
  \nonumber \\ &&
  - 2\,\mathcal{C}_{4}^{0,{\rm id.}q_1}
  - 2\,\mathcal{C}_{4}^{0,{\rm id.}q_3}
  + \delta(1-z) V^{(2)}_q|_{1/N}\,, \\
  \frac{1}{C_F N_F} \ff_{U,q}^{\PS,(2)} &=& 2\,\mathcal{B}_{4}^{0,{\rm id.}q'}\,,
\end{eqnarray}
for an identified quark and
\begin{align}
  \frac{1}{C_F N_F} \ff_{U,g}^{(2)}|_{N} &= 4\,\mathcal{A}_{4}^{0,{\rm id.}g}
  + 8\,\mathcal{A}_{3}^{1U,{\rm id.} g}\,, \\
  \frac{1}{C_F N_F} \ff_{U,g}^{(2)}|_{1/N} &= -2\,\tilde{\mathcal{A}}_{4}^{0,{\rm id.} g}
  -8\,\tilde{\mathcal{A}}_{3}^{1U,{\rm id.} g}\,,
\end{align}
for an identified gluon, respectively.
The notation is such that $X|_{Y}$ denotes the part of $X$ proportional to $Y$.
The superscript $U$ denotes the unrenormalised one-loop squared matrix elements,
see Section~\ref{sec:X31int}.
Finally, the $V^{(1)}_q$ and $V^{(2)}_q$ terms are given by
\begin{align}
  V^{(1)}_q &= \Re[\Delta(q^2,1)]\,F^{(1)}_{U,q}\,, \label{eq:Vq1} \\
  V^{(2)}_q &= \left(F^{(1)}_{U,q}\right)^2 + 2\,\Re[(\Delta(q^2,2)]\,F^{(2)}_{U,q}\,, \label{eq:Vq2}
\end{align}
where $F^{(1)}_{U,q}$ and $F^{(2)}_{U,q}$ are the unrenormalised first and
second order coefficients of the quark form factor respectively, with the
normalisation fixed by~\eqref{eq:ffexp} and by requiring that $F^{(0)}_{U,q}
= 1$, and
\begin{equation}
  \Delta(q^2,\kappa) = (-{\rm sgn}(q^2) - i0)^{-\kappa\e}\,.
\end{equation}
Explicit expressions for $F^{(1)}_{U,q}$ and $F^{(2)}_{U,q}$ can be found 
in~\cite{Matsuura:1988sm,Gehrmann:2005pd}.

In order to compare against the results of~\cite{Rijken:1996vr,Rijken:1996ns},
several convolutions between splitting functions and coefficient functions
needed to be computed; the Mathematica package
\textsc{MT}~\cite{Hoschele:2013pvt} has been extensively used.

\subsection{Identified hadrons in Higgs boson decay to gluons}
\label{sec:cfhiggs}

The gluon-gluon antenna functions $\mathcal{F}$, $\mathcal{G}$ and $\mathcal{H}$
have been derived~\cite{Gehrmann-DeRidder:2005alt} from the decay process
$H \to gg$ at NLO and NNLO, in the effective theory where the Higgs couples
directly to gluons, which is valid in the limit of infinitely massive quarks.
The second-order coefficient functions for the one-hadron inclusive Higgs decay
in such an effective theory were first obtained in~\cite{Almasy:2011eq}.
Here we are in the position to compare suitable combinations of integrated
antenna functions against the results of~\cite{Almasy:2011eq}, in a very
similar way to what we have done in Section~\ref{sec:cfgamma} for the
$\gamma^*/Z \to q\bar{q}$ decay.

Since the Higgs boson is a scalar, the parton fragmentation functions have only
one component, $\ft_i$, where a quark ($i=q$) or a gluon ($i=g$) is identified.
$\ft_i$ admits an expansion similar to~\eqref{eq:ffexp},
\begin{equation}\label{eq:ftexp}
  \ft_i = \ft_i^{(0)}
  + \left(\frac{\alpha_s}{4\pi} \right) \left( \frac{\mu_R^2}{Q^2} \right)^\e  \ft_i^{(1)}
  + \left(\frac{\alpha_s}{4\pi} \right)^2 \left( \frac{\mu_R^2}{Q^2} \right)^{2\e} \ft_i^{(2)}
  + \mathcal{O}(\alpha_s^3)\,,
\end{equation}
where we choose to adopt an expansion in terms of the renormalised coupling
constant $\alpha_s$, with $\mu_R^2$ the renormalisation scale, for ease of
comparison against~\cite{Almasy:2011eq}.
The normalisation is such that
\begin{equation}
  \ft_q^{(0)} = 0\,,\quad \ft_g^{(0)} = \delta(1-z)\,.
\end{equation}
At NLO we find the following relationships:
\begin{align}
  \frac{1}{T_F} \ft_q^{(1)} &= 8\,\mathcal{G}_{3}^{0,{\rm id.}q}\,, \\
  \ft_g^{(1)}|_N &= 2\,\mathcal{F}_{3}^{0,{\rm id.}g} + 4\,\delta(1-z) V^{(1)}_g|_{N}\,, \\
  \ft_g^{(1)}|_{N_F} &= 2\,\mathcal{G}_{3}^{0,{\rm id.}g} + 4\,\delta(1-z) V^{(1)}_g|_{N_F}\,,
\end{align}
whereas at NNLO we obtain 
\begin{align}
  \frac{}{T_F} \ft_q^{(2)}|_{N} &= 4\,\mathcal{G}_{4}^{0,{\rm id.}q}
  + 16\,\mathcal{G}_{3}^{1R,{\rm id.}q} \,,\\
  \frac{1}{T_F} \ft_q^{(2)}|_{1/N} &= -2\,\tilde{\mathcal{G}}_{4}^{0,{\rm id.}q}
  -16\,\tilde{\mathcal{G}}_{3}^{1R,{\rm id.} q}
  + 4\tilde{\mathcal{J}}_{4}^{0,{\rm id.}q}\,, \label{eq:FTq2invN} \\
  \frac{1}{T_F} \ft_q^{(2)}|_{N_F} &= 4\,\mathcal{H}_{4}^{0,{\rm id.} q}
  + 16\,\hat{\mathcal{G}}_{3}^{1R,{\rm id.} q}\,,
\end{align}
for the quark fragmentation function and 
\begin{align*}
  \ft^{(2)}_{g}|_{N^2} &= \mathcal{F}_{4}^{0,{\rm id.}g}
  + 4\,\mathcal{F}_{3}^{1R,{\rm id.}g} + 4\,\delta(1-z) V^{(2)}_g|_{N^2}\,, \\
  \ft^{(2)}_{g}|_{N N_F} &= 2\,\mathcal{G}_{4}^{0,{\rm id.}g}
  + 4\,\mathcal{G}_{3}^{1R,{\rm id.}g}
  + 4\,\hat{\mathcal{F}}_{3}^{1R,{\rm id.}g}
  + 4\,\delta(1-z) V^{(2)}_g|_{N N_F}\,, \\
  \ft^{(2)}_{g}|_{N_F/N} &= -\tilde{\mathcal{G}}_{4}^{0,{\rm id.}g}
  - 4\,\tilde{\mathcal{G}}_{3}^{1R,{\rm id.} g}
  + 4\,\delta(1-z) V^{(2)}_g|_{N_F/N}\,, \\
  \ft^{(2)}_{g}|_{N_F^2} &= 4\,\hat{\mathcal{G}}_{3}^{1R,{\rm id.}g}
  + 4\,\delta(1-z) V^{(2)}_g|_{N_F^2}\,, 
\end{align*}
for the gluon fragmentation function, respectively.
The superscript $R$ denotes the renormalised one-loop squared matrix elements,
i.e.\ the integrated one-loop three-particle antenna functions before the
subtraction of the integrated tree-level antenna function multiplied with the
virtual one-loop correction to the hard radiator vertex, see
Section~\ref{sec:X31int}.
$V^{(1)}_g$ and $V^{(2)}_g$ are related to the gluon form factor, and they are
defined in full analogy with the quark case (see~\eqref{eq:Vq1} and~\eqref{eq:Vq2}, respectively).
Finally, the antenna function $\tilde{\mathcal{J}}_{4}^{0,{\rm id.}q}$, which
appears in~\eqref{eq:FTq2invN}, comes from the infrared-finite interference of four quark
final states with identical quark flavour (see~\cite{Daleo:2009yj}). Its
expression is:
\begin{equation}
  \tilde{\mathcal{J}}_{4}^{0,{\rm id.}q} =
  \left( - \frac{9}{4} + \frac{7}{2} z - \frac{7}{2} z^2
    + \left( 2 - 4z + 4z^2 \right) \zeta_3  \right) + \mathcal{O}(\e)\,.
\end{equation}
As it is finite in all limits, it does not appear in Table~\ref{tab:X40}.

\section{Infrared structure of $e^+e^- \to \text{3 jets}$ with fragmentation at NLO}
\label{sec:epemNLO}

As an example of the formalism, in this section we present explicit expressions
for the antenna subtraction terms for one identified hadron in the $e^+e^- \to
\text{3 jets}$ process, as studied by OPAL~\cite{OPAL:2004prv}, at NLO. We keep
the notation as close as possible to~\cite{Gehrmann-DeRidder:2007foh}, where the
subtraction terms for the $e^+e^- \to \text{3 jets}$ process without
fragmentation can be found.

The short-distance cross sections with one identified parton $p=q,g,\bar{q}$ for
three-jet production at the leading order are given by:
\begin{eqnarray}
\der\hsig^{\B}_q &=& 
 N_{3} \,\der\Phi_{3}(p_{1},p_2,p_{3};\q) \,
 A^{0}_{3}(\id{1}_{q},3_{g},2_{\bar{q}})\, \JET_{3}^{(3)}(\{p_{1},p_2,p_{3}\};\eta p_1)\,,
 \label{eq:sigBq} \\
\der\hsig^{\B}_{\bar{q}} &=& 
 N_{3} \,\der\Phi_{3}(p_{1},p_2,p_{3};\q) \,
 A^{0}_{3}(1_{q},3_{g},\id{2}_{\bar{q}})\, \JET_{3}^{(3)}(\{p_{1},p_2,p_{3}\};\eta p_2)\,, 
 \label{eq:sigBqb} \\
\der\hsig^{\B}_g &=& 
 N_{3} \,\der\Phi_{3}(p_{1},p_2,p_{3};\q) \,
 A^{0}_{3}(1_{q},\id{3}_{g},2_{\bar{q}})\, \JET_{3}^{(3)}(\{p_{1},p_2,p_{3}\};\eta p_3)\,,
 \label{eq:sigBg} 
\end{eqnarray}
with $N_3$ a normalisation factor and $A^{0}_{3}$ the tree-level matrix element
squared (which coincides with the antenna function denoted with the same symbol).
For ease of readability, we denote the identified particle with a superscript
(id.), even though it is also indicated explicitly as argument in the jet
function.
We consider the NLO corrections to eqs.~\eqref{eq:sigBq}-\eqref{eq:sigBg} by
assuming the flavour of $q$ as fixed. Any other quark flavour different from $q$
will be denoted as $\qp$. At the end we can get the full NLO cross section for
$e^+e^- \to \text{3 jets}$ by summing over all possible tree-level quark
flavours.

\subsection{Real level}

At NLO, we can have three different four-parton final states: $q\qb
gg$, $q\qb\qp\qbp$ (non-identical quarks) and $q\qb q\qb$ (identical quarks).
The four-parton real radiation contribution to the NLO cross section when the
quark $q$ is identified is
\begin{eqnarray}
\der\hsig_{q}^R &=& \Bigg\{ \Bigg[ \frac{N}{2} \left(
\,A_{4}^0 (\id{1}_q,3_g,4_g,2_{\bar q})
+ A_{4}^0 (\id{1}_q,4_g,3_g,2_{\bar q}) \right)
-\frac{1}{2N}
\tilde A_{4}^0 (\id{1}_q,3_g,4_g,2_{\bar q})
\nonumber \\ && \phantom{\Bigg\{\Bigg[ }
+ B_{4}^0 (\id{1}_q,3_{q},4_{\bar{q}},2_{\bar q})
+ N_{F1} B_{4}^0 (\id{1}_q,3_{q'},4_{\bar{q}'},2_{\bar q})
\nonumber \\ && \phantom{\Bigg\{\Bigg[ }
- \frac{1}{N} \, \left(
C_{4}^0 (\id{1}_q,3_q,4_{\bar q},2_{\bar q})
+C_{4}^0 (2_{\bar q},4_{\bar q},3_q,\id{1}_q) \right) \Bigg]
\JET_{3}^{(4)}(\{p_{1},\ldots,p_{4}\};\eta p_1)
\nonumber \\ && \phantom{\Bigg\{}
+ \Bigg[
B_{4}^0 (1_q,\id{3}_{q},4_{\bar{q}},2_{\bar q})
- \frac{1}{N} \, \left(
C_{4}^0 (1_q,\id{3}_q,4_{\bar q},2_{\bar q})
+C_{4}^0 (2_{\bar q},4_{\bar q},\id{3}_q,1_q) \right) \Bigg]
\nonumber \\ && \phantom{\Bigg\{ +}
\times \JET_{3}^{(4)}(\{p_{1},\ldots,p_{4}\};\eta p_3) 
\Bigg\} N_{{4}}\,  \der\Phi_{4}(p_{1},\ldots,p_{4};\q) \,,
\label{eq:sigRq}
\end{eqnarray}
whereas the contribution when $q'$ is identified is given by
\begin{equation}
\der\hsig_{q'}^R = \Bigg\{ 
B_{4}^0 (1_q,\id{3}_{q'},4_{\bar{q}'},2_{\bar q})
\JET_{3}^{(4)}(\{p_{1},\ldots,p_{4}\};\eta p_3) \Bigg\}
N_{4}\,\der\Phi_{4}(p_{1},\ldots,p_{4};\q)\,.
\label{eq:sigRqp}
\end{equation}
The contributions when the anti-quark $\bar{q}$ or $\bar{q}'$ are identified are
similar to~\eqref{eq:sigRq} and~\eqref{eq:sigRqp}, respectively.
Finally, the contribution with a gluon identified reads
\begin{eqnarray}
\der\hsig_{g}^R &=& \Bigg\{
\sum_{(i,j) \in P(3,4)} \left[ \frac{N}{2} \left(
\,A_{4}^0 (1_q,\id{i}_g,j_g,2_{\bar q})
+ A_{4}^0 (1_q,j_g,\id{i}_g,2_{\bar q}) \right)
-\frac{1}{2N}
\tilde A_{4}^0 (1_q,\id{i}_g,j_g,2_{\bar q}) \right]
\nonumber \\ && \phantom{\times \Bigg\{}
\times \JET_{3}^{(4)}(\{p_{1},\ldots,p_{4}\}; \eta p_i) 
\Bigg\} N_{{4}}\,  \der\Phi_{4}(p_{1},\ldots,p_{4};\q) \,.
\label{eq:sigRg}
\end{eqnarray}

The subtraction terms for the contribution when the quark $q$ is identified reads
\begin{equation}
  \der\hsig_{q}^S = \der\hsig_{q(q)}^S + \der\hsig_{g(q)}^S\,,
\end{equation}
where in the subtraction term $\der\hsig_{q(q)}^S$ the mapped particle in
the reduced matrix element has the same flavour as the identified particle,
whereas in the subtraction term $\der\hsig_{g(q)}^S$ the the flavour of the
parton in the reduced matrix element is different from the flavour of the
particle in the antenna, called identity changing (IC) term.
Explicit expressions for the two subtraction terms are
\begin{eqnarray}
\der\hsig_{q(q)}^S&=& 
N_{{4}}\,  \der\Phi_{4}(p_{1},\ldots,p_{4};\q) 
\nonumber \\ && \times \Bigg\{ \sum_{(i,j) \in P(3,4)} \bigg[
\frac{N}{2}\, d_3^0(\id{1}_q,i_g,j_g)\, 
A_3^0(\widetilde{1}_q,\widetilde{(ij)}_g,2_{\bar q})\, 
{J}_{3}^{(3)}(\{\widetilde{1}_q,\widetilde{(ij)}_g,2_{\bar q}\};\eta p_1)
\nonumber \\ && \phantom{\times \Bigg\{ \sum_{(i,j) \in P(3,4)} \bigg[}
+ \frac{N}{2}\, d_3^0(2_{\bar q},i_g,j_g) \,
A_3^0(\id{1}_{q},\widetilde{(ji)}_g,\widetilde{(2i)}_{\bar q}) \,
{J}_{3}^{(3)}(\{1_{q},\widetilde{(ji)}_g,\widetilde{(2i)}_{\bar q}\};\eta p_1)
\nonumber \\ && \phantom{\times \Bigg\{ \sum_{(i,j) \in P(3,4)} \bigg[}
- \frac{1}{2N}\,  A_3^0(\id{1}_q,i_g,2_{\bar q}) \,
A_3^0(\widetilde{1}_q,j_g,\widetilde{(2i)}_{\bar q}) \,
{J}_{3}^{(3)}(\{\widetilde{1}_q,j_g,\widetilde{(2i)}_{\bar q}\};\eta p_1) \bigg]
\nonumber \\ && \phantom{\times \Bigg\{}
+ N_{F} \bigg[ E_3^0(\id{1}_q,3_{q'},4_{\bar q'})\, 
A_3^0(\widetilde{1}_q,\widetilde{(34)}_g,2_{\bar q})\, 
{J}_{3}^{(3)}(\{\widetilde{1}_q,\widetilde{(34)}_g,2_{\bar q}\}; \eta p_1)
\nonumber \\ && \phantom{\times \Bigg\{ + N_{F} \bigg[}
+  E_3^0(2_{\bar q},3_{q'},4_{\bar q'}) \,
A_3^0(\id{1}_{q},\widetilde{(34)}_g,\widetilde{(24)}_{\bar q}) \,
{J}_{3}^{(3)}(\{1_{q},\widetilde{(34)}_g,\widetilde{(24)}_{\bar q}\}; \eta p_1)
\bigg]\Bigg\}\,, \nonumber \\
\label{eq:sigSqq}
\end{eqnarray}
and
\begin{eqnarray}
\der\hsig_{g(q)}^S&=&
N_{{4}}\,  \der\Phi_{4}(p_{1},\ldots,p_{4};\q)
\nonumber \\ && \times
\bigg[ E_3^0(1_q,\id{3}_{q},4_{\bar q})\, 
A_3^0(\widetilde{(14)}_q,\widetilde{3}_g,2_{\bar q})\, 
{J}_{3}^{(3)}(\{\widetilde{(14)}_q,\widetilde{3}_g,2_{\bar q}\}; \eta p_3)
\bigg]\,.
\label{eq:sigSgq}
\end{eqnarray}
In eqs.~\eqref{eq:sigSqq}-\eqref{eq:sigSgq} we have denoted the mapped momenta
in the reduced matrix element in two different ways, according to the type of
mapping we are applying.  Given three particles $i$,$j$ and $k$, when none of
them is undergoing fragmentation, we adopt the standard NLO final-final mapping,
indicated as $\widetilde{(ij)}$ and $\widetilde{(jk)}$ e.g.\ in the second line of~\eqref{eq:sigSqq}; when one of them is fragmenting, say $i$, we use the NLO
fragmentation mapping of~\eqref{eq:NLOmap}, indicated as $\widetilde{i}$ and
$\widetilde{(jk)}$ e.g.\ in the first line of~\eqref{eq:sigSqq}.
The subtraction terms for the contribution when the quark $q'$ is purely
identity-changing (since there is no $q'$ at the Born level), and has the same
structure as~\eqref{eq:sigSgq}:
\begin{eqnarray}
\der\hsig_{q'} \equiv \der\hsig_{g(q')}^S&=& 
N_{{4}}\,  \der\Phi_{4}(p_{1},\ldots,p_{4};\q) 
\nonumber \\ && \times 
\bigg[ E_3^0(1_q,\id{3}_{q'},4_{\bar q'})\, 
A_3^0(\widetilde{(14)}_q,\widetilde{3}_g,2_{\bar q})\, 
{J}_{3}^{(3)}(\{\widetilde{(14)}_q,\widetilde{3}_g,2_{\bar q}\}; \eta p_3)
\bigg]\,. \nonumber \\
\end{eqnarray}
Notice that, since the antenna function $E_3^0$ contains only the
$3 \parallel 4$ collinear limit, we are free to choose as third momentum in the
antenna either the particle 1 or 2.

Finally, the subtraction term for the contribution where the gluon is identified
is given by the sum of three contributions
\begin{equation}
  \der\hsig_{g}^S =
  \der\hsig_{g(g)}^S + \der\hsig_{q(g)}^S + \der\hsig_{\qb(g)}^S\,,
\end{equation}
where
\begin{eqnarray}
\lefteqn{\der\hsig_{g(g)}^S= N_{{4}}\,  \der\Phi_{4}(p_{1},\ldots,p_{4};\q)}
\nonumber \\ && \times \sum_{(i,j) \in P(3,4)} \Bigg\{ 
\frac{N}{2}\, d_{3,g\to g}^0(1_q,j_g,\id{i}_g)\, 
A_3^0(\widetilde{(1j)}_q,\widetilde{i}_g,2_{\bar q})\, 
{J}_{3}^{(3)}(\{\widetilde{(1j)}_q,\widetilde{i}_g,2_{\bar q}\}; \eta p_i)
\nonumber \\ && \phantom{\times \sum_{(i,j) \in P(3,4)} \Bigg\{}
+ \frac{N}{2}\, d_{3,g\to g}^0(2_{\bar q},j_g,\id{i}_g) \,
A_3^0(1_{q},\widetilde{i}_g,\widetilde{(2j)}_{\bar q}) \,
{J}_{3}^{(3)}(\{1_{q},\widetilde{i}_g,\widetilde{(2j)}_{\bar q}\}; \eta p_i)
\nonumber \\ && \phantom{\times \sum_{(i,j) \in P(3,4)} \Bigg\{}
- \frac{1}{2N}\,  A_3^0(1_{q},j_g,2_{\bar{q}}) \,
A_3^0(\widetilde{(1j)}_q,\id{i}_g,\widetilde{(2j)}_{\bar q}) \,
{J}_{3}^{(3)}(\{\widetilde{(1j)}_q,i_g,\widetilde{(2j)}_{\bar q}\}; \eta p_i)
\bigg] \Bigg\}\,, \nonumber \\
\label{eq:sigSgg}
\end{eqnarray}
\begin{eqnarray}
\der\hsig_{q(g)}^S&=& N_{{4}}\,  \der\Phi_{4}(p_{1},\ldots,p_{4};\q)
\nonumber \\ && \times \sum_{(i,j) \in P(3,4)} \Bigg\{ 
\frac{N}{2}\, d_{3,q\to g}^0(1_q,\id{i}_g,j_g)\, 
A_3^0(\widetilde{i}_q,\widetilde{(1j)}_g,2_{\bar q})\, 
{J}_{3}^{(3)}(\{\widetilde{i}_q,\widetilde{(1j)}_g,2_{\bar q}\}; \eta p_i)
\nonumber \\ && \phantom{\times \sum_{(i,j) \in P(3,4)} \Bigg\{}
- \frac{1}{2N}\, d_{3,q\to g}^0(1_q,\id{i}_g,j_g) \,
A_3^0(\widetilde{i}_q,\widetilde{(1j)}_g,2_{\bar q})\, 
{J}_{3}^{(3)}(\{\widetilde{i}_q,\widetilde{(1j)}_g,2_{\bar q}\}; \eta p_i)
\bigg] \Bigg\}\,, \nonumber \\
\label{eq:sigSqg}
\end{eqnarray}
and
\begin{eqnarray}
\der\hsig_{\qb(g)}^S&=& N_{{4}}\,  \der\Phi_{4}(p_{1},\ldots,p_{4};\q)
\nonumber \\ && \times \sum_{(i,j) \in P(3,4)} \Bigg\{ 
\frac{N}{2}\, d_{3,q\to g}^0(2_{\bar q},\id{i}_g,j_g) \,
A_3^0(1_{q},\widetilde{(2j)}_g,\widetilde{i}_{\bar q}) \,
{J}_{3}^{(3)}(\{1_{q},\widetilde{(2j)}_g,\widetilde{i}_{\bar q}\}; \eta p_i)
\nonumber \\ && \phantom{\times \sum_{(i,j) \in P(3,4)} \Bigg\{}
- \frac{1}{2N}\,  d_{3,q\to g}^0(2_{\bar{q}},\id{i}_g,j_{g}) \,
A_3^0(1_{q},\widetilde{(2j)}_g,\widetilde{i}_{\bar q}) \,
{J}_{3}^{(3)}(\{1_{q},\widetilde{(2j)}_g,\widetilde{i}_{\bar q}\}; \eta p_i)
\bigg] \Bigg\}\,. \nonumber \\
\label{eq:sigSqbg}
\end{eqnarray}
Note that in eqs.~\eqref{eq:sigSqg}-\eqref{eq:sigSqbg}, in order to remove the
quark-gluon collinear divergence in the photon-like matrix element $\tilde
A_{4}^0$, we have used the sub-antenna $d_{3,q\to g}^0(k_q,\id{i}_g,j_j)$, which
contains only the collinear limit $i \parallel k$ (see comment after~\eqref{eq:splitD30}).

\subsection{Virtual level}

The virtual one-loop contribution to $\gamma^*\to q\bar q g$ is
\begin{eqnarray}
\der\hsig^V &=& 
N_{{3}}\,\der\Phi_{3}(p_{1},\ldots,p_{3};\q)\, {J}_{3}^{(3)}(p_1,p_2,p_3) 
\nonumber \\
&& \times
\, \bigg( N \left[
A_3^1(1_q,3_g,2_{\bar q})
+{\cal A}_2^1(s_{123}) A_3^0(1_q,3_g,2_{\bar q})\right]\nonumber \\
&& \phantom{\times \,\bigg(}
-\frac{1}{N}  \left[
\tilde
A_3^1(1_q,3_g,2_{\bar q})
+{\cal A}_2^1(s_{123}) A_3^0(1_q,3_g,2_{\bar q})\right]
+N_F \hat A_3^1(1_q,3_g,2_{\bar q})\bigg)\, ,
\end{eqnarray}
where any of the three particles can be identified, thus generating the three
terms $\der\hsig^V_q$, $\der\hsig^V_{\qb}$ and $\der\hsig^V_{g}$.
The infrared behaviour of the virtual matrix elements reads
\begin{eqnarray}
  {\rm Poles}(A_3^1(1_q,3_g,2_{\bar q})) &=& 2 ( \mathbf{I}^{(1)}_{qg}(\e,s_{13})
  +  \mathbf{I}^{(1)}_{qg}(\e,s_{23}) -  \mathbf{I}^{(1)}_{q\bar{q}}(\e,s_{123}) ) A_3^0(1,3,2)\,, \\
 {\rm Poles}(\tilde{A}_3^1(1_q,3_g,2_{\bar q})) &=& 2 ( \mathbf{I}^{(1)}_{q\bar{q}}(\e,s_{12})
  -  \mathbf{I}^{(1)}_{q\bar{q}}(\e,s_{123}) ) A_3^0(1,3,2)\,, \\
 {\rm Poles}(\hat{A}_3^1(1_q,3_g,2_{\bar q})) &=& 2 ( \mathbf{I}^{(1)}_{qg,F}(\e,s_{13})
  +  \mathbf{I}^{(1)}_{qg,F}(\e,s_{23}) ) A_3^0(1,3,2)\,, \label{eq:polesA31h}\\
{\rm Poles}(\mathcal{A}_2^1(s_{123})) &=& 2 \mathbf{I}^{(1)}_{q\bar{q}}(\e,s_{123})\,,
\end{eqnarray}
where $\mathbf{I}^{(1)}_{xy}$ are the colour-ordered singularity operators,
whose expressions are reported in eqs.~\eqref{eq:IRopA}-\eqref{eq:IRopZ}.
The integral of the subtraction term $\der\hsig_q^{\S}$ reads:
\begin{eqnarray}
\int_1 \der\hsig_{q(q)}^S&=& 
N_{{3}}\,\der\Phi_{3}(p_{1},\ldots,p_{3};\q)\,{J}_{3}^{(3)}(\{p_1,p_2,p_3\}; \eta p_1)
\nonumber \\ && \times \bigg[
N\,\left(
\frac{1}{2} {\cal D}_3^{0,{\rm id.}q}(s_{13},z)
+ \frac{1}{2} {\cal D}_3^0(s_{23})\right) - \frac{1}{N} 
{\cal A}_3^{0,{\rm id.}q}(s_{12},z)
\nonumber \\ && \phantom{\times\Bigg\{\bigg[} 
+ N_F \,\left(  
{\cal E}_3^{0,{\rm id.}q}(s_{13},z)+{\cal E}_3^0(s_{23})\right)\bigg]
A_3^0(\id{1}_{q},3_g,2_{\bar q}) \,, 
\label{eq:intsigSqq}  
\end{eqnarray}
\begin{eqnarray}
\int_1 \der\hsig_{g(q)}^S &=&
N_{{3}}\,\der\Phi_{3}(p_{1},\ldots,p_{3};\q)\,{J}_{3}^{(3)}(\{p_1,p_2,p_3\}; \eta p_3)
\nonumber \\ && \times\,
{\cal E}_3^{0,{\rm id.}q'}(s_{13},z)
A_3^0(1_{q},\id{3}_g,2_{\bar q}) \,,
\end{eqnarray}
where in~\eqref{eq:intsigSqq} we need the integral of the inclusive sub-antenna
$d_3^0$, which is given by one-half the integral ${\cal D}_3^0$ of the full inclusive
antenna $D_3^0$.
The integral of the subtraction term $\der\hsig_g^{\S}$ reads:
\begin{eqnarray}
\int_1 \der\hsig_{g(g)}^S&=& 
N_{{3}}\,\der\Phi_{3}(p_{1},\ldots,p_{3};\q)\,{J}_{3}^{(3)}(\{p_1,p_2,p_3\}; \eta p_3)
\nonumber \\ && \times \Bigg\{ N \,\Bigg[
\left( {\cal D}_{3,g\to g}^{0,{\rm id.}g}(s_{13},z)+{\cal D}_{3,g\to g}^{0,{\rm id.}g}(s_{23},z) \right)
  \Bigg]  - \frac{1}{N}\,\Bigg[ {\cal A}_3^{0}(s_{12}) \Bigg]
\Bigg\} A_3^0(1_{q},\id{3}_g,2_{\bar q})  \,, \nonumber \\
\end{eqnarray}
\begin{eqnarray}
\int_1 \der\hsig_{q(g)}^S&=& 
N_{{3}}\,\der\Phi_{3}(p_{1},\ldots,p_{3};\q)\,{J}_{3}^{(3)}(\{p_1,p_2,p_3\}; \eta p_1) 
\nonumber \\ && \times \Bigg\{ N \,\Bigg[ {\cal D}_{3,q\to g}^{0,{\rm id.}g}(s_{13},z) \Bigg]
- \frac{1}{N}\,\Bigg[ {\cal D}_{3,q\to g}^{0,{\rm id.}g}(s_{13},z) \Bigg] 
\Bigg\} A_3^0(\id{1}_{q},3_g,2_{\bar q}) \,.
\label{eq:intsigSqg}
\end{eqnarray}
The integral of $\der\hsig_{\qb(g)}^S$ is given by~\eqref{eq:intsigSqg} with
$1 \leftrightarrow 2$.
Finally the integral of $\der\hsig_{g(q')}^S$ is
\begin{equation}
  \int_1 \der\hsig_{g(q')}^S = \int_1 \der\hsig_{g(q)}^S\,.
\end{equation}

Given the pole structure of the fully integrated antenna
functions~\cite{GehrmannDeRidder:2005cm},
\begin{eqnarray}
  {\rm Poles}({\cal D}_3^0(q^2)) &=& -4\mathbf{I}^{(1)}_{qg}(\e,q^2)\,, \nonumber \\
  {\rm Poles}({\cal A}_3^0(q^2)) &=& -2\mathbf{I}^{(1)}_{q\bar{q}}(\e,q^2)\,, \nonumber \\
  {\rm Poles}({\cal E}_3^0(q^2)) &=& -4\mathbf{I}^{(1)}_{qg,F}(\e,q^2)\,,
\end{eqnarray}
and the explicit expressions of the integrated fragmentation antenna functions
$\XTZ{X}{p}$ provided in Appendix~\ref{app:X30int}, we see that most of the poles
cancel, except the ones proportional to splitting functions, which are removed
by means of the mass factorisation counterterms, such that
\begin{eqnarray}
  {\rm Poles}\left(\der\hsig^V_g + \int_1 \der\hsig_{g}^S + \der\hsig^{\MF}_g\right) &=& 0\,,
  \label{eq:polesg}\\
  {\rm Poles}\left(\der\hsig^V_q + \int_1 \der\hsig_{q}^S + \der\hsig^{\MF}_q\right) &=& 0\,, \\
  {\rm Poles}\left(\int_1 \der\hsig_{\qp}^S + \der\hsig^{\MF}_{\qp}\right) &=& 0\,,
\end{eqnarray}
thus yielding an infrared-finite result.
Moreover, note that there is a one-to-one correspondence between
$\int \der\hsig_{i(j)}$ and $\MFKB{1}{j \rto i}$ inside $\der\hsig^{\MF}_j$, in
the sense that the poles proportional to splitting kernels in the former are
explicitly removed by the latter. In particular, the integral of IC subtraction terms such as 
$\der\hsig^S_{g(\qp)}$ does not have a corresponding virtual contributions and its infrared poles are removed
entirely by the mass factorisation term.

It is instructive to look explicitly at the poles of~\eqref{eq:polesg},
before adding the mass factorisation term:
\begin{eqnarray}
  && {\rm Poles}\left( \der\hsig^V_g + \int_1 \der\hsig_{g}^S \right)
  \nonumber \\ && \quad = N_{{3}}\,\int \der z\, N \Bigg\{
  \Bigg[ (s_{13})^{-\e} + (s_{23})^{-\e} \Bigg]
  \bigg( -\frac{1}{2\e} p^{(0)}_{gg}(z) \bigg)\,\der\hsig^{\B}_{g}(\eta z)
  \nonumber \\ && \phantom{\quad = N_{{3}}\,\int \der z\, N \Bigg\{} +
  (s_{13})^{-\e} \bigg( -\frac{1}{2\e}p^{(0)}_{gq}(z) \bigg)\,\der\hsig^{\B}_{q}(\eta z) +
  (s_{23})^{-\e} \bigg( -\frac{1}{2\e}p^{(0)}_{gq}(z) \bigg)\,\der\hsig^{\B}_{\qb}(\eta z)
  \Bigg\}
  \nonumber \\ &&  \phantom{\quad = N_{{3}}\,\int \der z\,} - \frac{1}{N} \Bigg\{
  (s_{13})^{-\e} \bigg( -\frac{1}{2\e}p^{(0)}_{gq}(z) \bigg)\,\der\hsig^{\B}_{q}(\eta z)
  + (s_{23})^{-\e} \bigg( -\frac{1}{2\e}p^{(0)}_{gq}(z) \bigg)\,\der\hsig^{\B}_{\qb}(\eta z)
  \Bigg\}
  \nonumber \\ && \phantom{\quad = N_{{3}}\,\int \der z\,} + N_F \Bigg\{
  \Bigg[ \Re(-s_{13})^{-\e} + \Re(-s_{23})^{-\e} \Bigg] \frac{1}{6\e} \der\hsig^{\B}_{g}(\eta z)
  \Bigg\}
  \label{eq:polVmS}
\end{eqnarray}
The last line contains the pole coming from $\hat{A}_3^1$, whose singularity
structure~\eqref{eq:polesA31h} is given by twice the infrared operator
$\mathbf{I}^{(1)}_{qg,F}$~\eqref{eq:IRIqgF}. Since there is no term coming
from the integral of $\der\hsig_g^S$ proportional to $N_F$, such a pole is
entirely canceled by $\MFK{1}{gg,F}$.
Note that the $\e$-poles in~\eqref{eq:polVmS} appear together with different
invariants raised to the $(-\e)$ power i.e.\ differing by $\mathcal{O}(\e)$,
thus allowing for a cancellation of the poles, by leaving the usual logarithms
of ratio of scales as leftover when $\e \to 0$.

\section{Conclusions}
\label{sec:concl}

In this paper, we have described how identified final-state hadrons can be incorporated 
in the antenna subtraction formalism for NNLO calculations, which  required the 
introduction of fragmentation antenna functions.
These functions retain the information on 
a final-state parton momentum fraction, in contrast to previously considered 
antenna functions that were inclusive in the final state parton momenta. 

In the  description of the formalism, we focused on identified hadron production 
processes in generic hadronic final states in $e^+e^-$ annihilation. This 
restriction is largely for notational simplicity. The structure of the formalism, 
which amounts to the introduction of new subtraction terms for all 
unresolved configurations that involve the parton that subsequently fragments into the 
identified hadron, carries over to electron-hadron and hadron-hadron collisions,
as already demonstrated for identified photons~\cite{Gehrmann:2022cih} at hadron colliders. 

We have outlined the structure of the subtraction terms that are newly required
for unresolved configurations involving an identified final-state parton. In
$e^+e^-$ annihilation, these are constructed from antenna functions with both
radiator partons in the final state (final-final kinematics). We introduced
suitable phase space factorisations and mappings at NLO and NNLO, which retained
the dependence on the momentum fraction $z$ of the fragmenting parton.  The
relevant antenna functions have been integrated over the factorised phase space
by leaving $z$ unintegrated, in order to combine with mass factorisation terms
at the virtual, real-virtual and double-virtual level.
Since the antenna functions are related to physical matrix
elements~\cite{Gehrmann-DeRidder:2004ttg,Gehrmann-DeRidder:2005svg,Gehrmann-DeRidder:2005alt},
we have been able to check our integrated results against known expressions in
the literature for single-inclusive coefficient functions in vector boson and
Higgs decay.

The integrated antenna functions are inclusive over unresolved radiation, but in
the context of a subtraction scheme they can be used as local subtraction terms
for more {\em exclusive} calculations.
For instance, a NNLO calculation of the hadron-in-jet fragmentation process in
three-jet final states in $e^+e^-$ annihilation, whose NLO subtraction structure
has been detailed in Section~\ref{sec:epemNLO}, could be envisaged. Experimental
data differential in $x_E = E_h/E_{jet}$, where $E_h$ is the energy of the
hadron $h$ and $E_{jet}$ is the energy of the jet to which it is assigned, have
been published, see e.g.\ \cite{OPAL:2004prv}. In the latter paper, the
experimental data are compared with NLO calculations, and they fail to describe
the full set of results; a re-analysis of such $e^+e^-$ data at NNLO accuracy
would thus be warranted. Moreover, the hadron-in-jet data have been proven to
provide valuable constraints on fragmentation
functions~\cite{Kaufmann:2015hma,Anderle:2017cgl}.

The fragmentation antenna functions in final-final kinematics derived here will
also appear in the construction of subtraction terms for processes with
identified hadrons in deep-inelastic scattering or at hadron colliders.  In
these cases, one (but not two) of the hard radiators can be in the initial
state. The resulting fragmentation antenna functions in initial-final kinematics
were already derived in parts in the context of photon fragmentation up to
NNLO~\cite{Gehrmann:2022cih}, their completion for all parton combinations will
be addressed in future work.

\acknowledgments We are indebited to Robin Sch\"urmann for constructive inputs
and useful discussions during the course of this work. Discussions with Petr
Jakub$\check{{\rm c}}$ik, Matthias Kerner, Matteo Marcoli, and Tong-Zhi Yang are also gratefully
acknowledged.
This work has received funding from the Swiss National Science Foundation (SNF)
under contract 200020-204200 and from the European Research Council (ERC) under
the European Union's Horizon 2020 research and innovation programme grant
agreement 101019620 (ERC Advanced Grant TOPUP).

\appendix
\section{Time-like mass factorisation kernels}
\label{app:mfks}

Before factorisation of final-state mass singularities, the physical cross
section is written as the convolution of a {\em bare} fragmentation function
with the short-distance cross section still containing final-state collinear
divergences: symbolically,
\begin{equation}\label{eq:convbare}
  \der\hsig = D_i^B \conv \der\hsig_i^B\,.
\end{equation}
The bare fragmentation function is related to the physical mass-factorised
fragmentation function by means of
\begin{equation}\label{eq:Dp2b}
  D_i^B(z,\mu_a^2) = \sum_j D_j(z) \conv \mathbf{\Gamma}_{j \leftarrow i}(z,\mu_a^2) \,.
\end{equation}
where $\mathbf{\Gamma}_{j \leftarrow i}$ are the mass factorisation kernels, with a bold
letter to indicate that they carry colour factors, and $\mu_a^2$ is the
fragmentation scale.
The replacement $D_i^B \to D_i$ in~\eqref{eq:convbare} generates the mass
factorisation terms which are added to the short-distance cross
section. Symbolically, we then have:
\begin{equation}
  D_i^B \conv \der\hsig_i^B =  D_i(\mu_a^2) \conv \der\hsig_i(\mu_a^2)\,,
\end{equation}
where now on the right hand side each term retains a dependence on $\mu_a^2$, and
\begin{equation}
  \der\hsig_i(z,\mu_a^2) = \Big( \der\hsig_i(z)
  + \left(\frac{\as}{2\pi}\right) \der\hsig_i^{\MF,\NLO}(z, \mu_a^2)
  + \left(\frac{\as}{2\pi}\right)^2 \der\hsig_i^{\MF,\NNLO}(z, \mu_a^2) \Big)\,.
\end{equation}
The mass factorisation terms are obtained after an expansion of~\eqref{eq:Dp2b}, and they read
\begin{equation}
  \der\hsig_i^{\MF,\NLO} =
  - \overline{C}(\e) (\mu_a^2)^{-\e} \sum_j \mathbf{\Gamma}^{(1)}_{i \leftarrow j} \conv \der\hsig_j^{\LO}\,,
\end{equation}
and
\begin{equation}
  \der\hsig_i^{\MF,\NNLO} =
  - \overline{C}(\e)^2 (\mu_a^2)^{-2\e}
   \left[
   \sum_j \mathbf{\Gamma}^{(2)}_{i \leftarrow j} \conv \der\hsig_j^{\LO} 
   + \sum_j \mathbf{\Gamma}^{(1)}_{i \leftarrow j} \conv \der\hsig_j^{\NLO} 
  \right]\,.
\end{equation}
The mass factorisation kernels are given by:
\begin{equation}
  \mathbf{\Gamma}^{(1)}_{i \leftarrow j}  = -\frac{1}{\e}\,\mathbf{P}^{(0)}_{ij}\,,
\end{equation}
and 
\begin{equation}\label{eq:MF2}
  \mathbf{\Gamma}^{(2)}_{i \leftarrow j} =
  -\frac{1}{2\e} \mathbf{P}^{(1)}_{ij}
  +\frac{\beta_0}{2\e^2}\, \mathbf{P}^{(0)}_{ij}
  + \frac{1}{2\e^2} \left[ \mathbf{P}^{(0)}_{ik}
    \conv \mathbf{P}^{(0)}_{kj} \right]\,,
\end{equation}
where $\mathbf{P}^{(0)}$ and $\mathbf{P}^{(1)}$ are the leading
order~\cite{Altarelli:1977zs} and next-to-leading
order~\cite{Curci:1980uw,Furmanski:1980cm,Floratos:1981hs} time-like splitting
functions, respectively, whose expression can be found in many places, see e.g.\ \cite{Ellis:1996mzs}.
Time-like splitting functions coincide with the space-like splitting functions
at leading order, but they differ at higher orders.

Due to the intrinsic colour decomposition of the antenna functions, we need to
decompose the splitting kernels into the $N$ and $\NF$ factors, in order to
define colour stripped kernels to be used in the mass factorisation term, along
the lines of Appendix A of~\cite{Currie:2013vh}.
The one-loop mass factorisation kernels in the time-like region are the same as
the ones in the space-like region and are given by:
\begin{eqnarray}
  \MFKB{1}{q \rto q}(z) &=& \left(\frac{N^2-1}{N}\right)\,\MFK{1}{qq}(z)
  = \left(\frac{N^2-1}{N}\right)\,\left[-\frac{1}{2\e}\, p_{qq}^{(0)}(z)\right] \,, \\
  \MFKB{1}{g \rto q}(z) &=& \left(\frac{N^2-1}{N}\right)\,\MFK{1}{gq}(z)
  = \left(\frac{N^2-1}{N}\right)\,\left[-\frac{1}{2\e}\, p_{gq}^{(0)}(z)\right] \,, \\
  \MFKB{1}{q \rto g}(z) &=& \MFK{1}{qg}(z)
  = -\frac{1}{2\e}\, p_{qg}^{(0)}(z) \,, \\
  \MFKB{1}{g \rto g}(z) &=& N\,\MFK{1}{gg}(z) + N_F\,\MFK{1}{gg,F}(z)
  = N \left[ -\frac{1}{\e}\, p_{gg}^{(0)} \right]
  + N_F \left[ -\frac{1}{\e}\, p_{gg,F}^{(0)} \right]\,,
\end{eqnarray}
where we have exploited $C_F = (N^2-1)/(2 N)$ and $T_R = 1/2$. The set of
$p_{ij}^{(0)}$ reads:
\begin{eqnarray}
  p_{qq}^{(0)}(z) &=& \frac{3}{2}\delta(1-z) + 2\Dn{0}(1-z) - 1 - z\,, \\
  p_{gq}^{(0)}(z) &=& \frac{2}{z} - 2 + z\,,\\
  p_{qg}^{(0)}(z) &=& 1 - 2z + 2z^2\,,\\
  p_{gg}^{(0)}(z) &=& \frac{11}{6}\delta(1-z) + 2\Dn{0}(1-z) + \frac{2}{z} -4 +2z - 2z^2\,,\\
  p_{gg,F}^{(0)}(z) &=& -\frac{1}{3}\delta(1-z)\,.  
\end{eqnarray}
The two-loop mass factorisation kernels can be easily assembled according
to~\eqref{eq:MF2}; this is a rather straightforward procedure, so we do not
report here the resulting (lengthy) explicit expressions.

\section{Integrated NLO fragmentation antenna functions}
\label{app:X30int}

We report here the integrated form of the $X_3^0$ antenna functions, differential
in the momentum fraction $z$, as defined in~\eqref{eq:zdef}.
We recall the reader that the unintegrated $X_3^0$ antenna functions can be
found in~\cite{GehrmannDeRidder:2005cm}.
The discussion is similar to Appendix~B.2 of~\cite{Gehrmann:2022cih}.  However,
note that compared to~\cite{Gehrmann:2022cih}, we do not have a reference
particle and the definition of $z$ is different (compare~\eqref{eq:zdef}
with~(3.10) of~\cite{Gehrmann:2022cih}): this results in different integrated
antenna functions.

It is convenient to introduce the NLO colour-ordered singularity
operators~\cite{Gehrmann-DeRidder:2005btv}, which appear in the explicit
expressions of the integrated antenna functions:
\begin{eqnarray}
  \mathbf{I}_{q\bar{q}}^{(1)}(\e,s_{q\qb}) &=&
  -\frac{e^{\e\gamma}}{2\Gamma(1-\e)} \left[ \frac{1}{\e^2} + \frac{3}{2\e} \right]
  \Re(-s_{q\qb})^{-\e}\,, \label{eq:IRopA} \\
  \mathbf{I}_{qg}^{(1)}(\e,s_{qg}) &=&
  -\frac{e^{\e\gamma}}{2\Gamma(1-\e)} \left[ \frac{1}{\e^2} + \frac{5}{3\e} \right]
  \Re(-s_{qg})^{-\e}\,,\\
    \mathbf{I}_{gg}^{(1)}(\e,s_{gg}) &=&
  -\frac{e^{\e\gamma}}{2\Gamma(1-\e)} \left[ \frac{1}{\e^2} + \frac{11}{6\e} \right]
  \Re(-s_{gg})^{-\e}\,,\\
    \mathbf{I}_{q\bar{q},F}^{(1)}(\e,s_{q\qb}) &=& 0\,,\\
    \mathbf{I}_{qg,F}^{(1)}(\e,s_{qg}) &=&
  \frac{e^{\e\gamma}}{2\Gamma(1-\e)} \frac{1}{6\e} \Re(-s_{qg})^{-\e}\,, \label{eq:IRIqgF}\\
  \mathbf{I}_{gg,F}^{(1)}(\e,s_{gg}) &=&
  \frac{e^{\e\gamma}}{2\Gamma(1-\e)} \frac{1}{3\e} \Re(-s_{gg})^{-\e}\,. \label{eq:IRopZ}
\end{eqnarray}

At NLO, there is one quark-quark antenna function, $A_3^0(1_q, 2_g, 3_{\qb})$,
symmetric under exchange of the quark pair. When the quark or anti-quark is
identified, we obtain
\begin{eqnarray}
  \XTZ{A}{q}(q^2,z) &=& -2\mathbf{I}_{q\bar{q}}^{(1)}(\e,q^2)\delta(1-z)
  + (q^2)^{-\e}\bigg[-\frac{1}{2\e}p^{(0)}_{qq}(z)
    \nonumber \\ &&     
    + \delta(1-z) \left(\frac{7}{4} + \frac{\pi^2}{3}\right)
    -\frac{3}{4} \Dn{0}(1-z) + \Dn{1}(1-z)
    \nonumber \\ &&         
    + \log(z) \left(\frac{1+z^2}{1-z}\right)
    - \frac{1}{2} \log(1-z) \left(1 + z\right)
    +\frac{5}{4} -\frac{3}{4} z    
    \bigg]  + \mathcal{O}(\e)\,,
  \label{eq:intA30q}  
\end{eqnarray}
whereas when the gluon is identified we get
\begin{eqnarray}
  \XTZ{A}{g}(q^2,z) &=& (q^2)^{-\e}\bigg[-\frac{1}{\e}p^{(0)}_{gq}(z)
    + \log(z) \left( \frac{4}{z} + 2z-4 \right)
    \nonumber \\ &&         
    - \log(1-z) \left( -\frac{2}{z} -z +2 \right)
    \bigg]  + \mathcal{O}(\e)\,.
  \label{eq:intA30g}
\end{eqnarray}
In~\eqref{eq:intA30g}, there is no infrared singularity operator associated
to the $q\qb$ vertex (as in~\eqref{eq:intA30q}), because the gluon has to be
resolved in order to be identified.

There are two quark-gluon antenna functions, $D_3^0(1_q,2_g,3_g)$ and
$E_3^0(1_q,2_{\qp},3_{\qbp})$.
The $D_3^0$ antenna function can be written as a sum of two sub-antenna
functions~\cite{Daleo:2006xa}:
\begin{equation}\label{eq:splitD30}
  D_3^0(1_q,2_g,3_g) = d_{3,q\to g}^0(1_q, 2_g, 3_g) + d_{3,g\to g}^0(1_q, 3_g, 2_g)\,,
\end{equation}
where the sub-antenna $d_{3,q\to g}^0(1,2,3)$ contains the collinear $1 \parallel 2$
limit, but not the collinear $1 \parallel 3$ limit nor the soft $3$ limit; it is
very handy in subtracting one single collinear limit --- see for instance its
usage in the NLO subtraction terms of Section~\ref{sec:epemNLO}.
Instead, the sub-antenna $d_{3,g\to g}^0(1,3,2)$ is singular in $s_{13}$ and
$s_{23}$, but not in $s_{12}$.
When we identify the quark, we integrate the full antenna $D_3^0$, to find
\begin{eqnarray}
  \XTZ{D}{q}(q^2,z) &=& -4\mathbf{I}^{(1)}_{qg}(\e,q^2)\delta(1-z)
  + (q^2)^{-\e}\bigg[-\frac{1}{\e}p^{(0)}_{qq}(z)
    + \delta(1-z)\left(\frac{67}{18}+\frac{2\pi^2}{3}\right)
    \nonumber \\ &&    
    -\frac{11}{6} \Dn{0}(1-z)
    + 2\Dn{1}(1-z)
    + \log(z) \left(-2+\frac{4}{1-z}-2z \right)
    \nonumber \\ &&    
    - \log(1-z) (1+z)
    +\frac{17}{6}+\frac{5}{6}z+\frac{1}{3}z^2
    \bigg]   + \mathcal{O}(\e)\,.
\end{eqnarray}
Instead, when we identify the gluon, we integrate the sub-antenna functions
introduced in~\eqref{eq:splitD30}. In both cases, we identify the gluon 2.
If we integrate $d_{3,q\to g}^0$ we obtain
\begin{eqnarray}
  \XTZd{D}{g}{q\to g}(q^2,z) &=&  
  (q^2)^{-\e} \bigg[ -\frac{1}{2\e}p^{(0)}_{gq}(z)
    + \log(z)\left(-2+\frac{2}{z}+z\right)
    \nonumber \\ &&            
    - \log(1-z) \left(1-\frac{1}{z}-\frac{1}{2}z\right)
    -z + \frac{3}{4}z^2
    \bigg]   + \mathcal{O}(\e)\,,
  \label{eq:intD30gq}  
\end{eqnarray}
whereas if we integrate $d_{3,g\to g}^0$ we get
\begin{eqnarray}
  \XTZd{D}{g}{g\to g}(q^2,z) &=&
  -2\mathbf{I}^{(1)}_{qg}(\e,q^2)\delta(1-z)        
  + (q^2)^{-\e} \bigg[ -\frac{1}{2\e}p^{(0)}_{gg}(z)
    + \delta(1-z) \left(\frac{7}{4} + \frac{\pi^2}{3}\right)
    \nonumber \\ &&            
    -\frac{3}{4} \Dn{0}(1-z)
    + \Dn{1}(1-z)
    + \log(z) \left(-4+\frac{2}{z} + \frac{2}{1-z}+2z-2z^2\right)
    \nonumber \\ &&            
    -\log(1-z) \left( 2 - \frac{1}{z} - z + z^2 \right)
    + \frac{3}{4} + \frac{9}{4} z
    \bigg]   + \mathcal{O}(\e)\,.
  \label{eq:intD30gg}
\end{eqnarray}
The pattern emerging from~\eqref{eq:intD30gq} and~\eqref{eq:intD30gg} is
interesting. Eq.~\eqref{eq:intD30gq} contains only a single pole proportional to
the splitting kernel $p_{gq}^{(0)}$, related to the $q \to qg$
branching. Eq.~\eqref{eq:intD30gg}, instead, has a richer structure, with the
presence of the infrared singularity operator $\mathbf{I}^{(1)}_{qg}$ (encoding
the virtual correction to a $qg$ vertex) and the splitting kernel $p_{gg}^{(0)}$
(encoding the outgoing gluon splitting into a pair of gluons).

The other quark-gluon antenna function, $E_3^0(1_q,2_{\qp},3_{\qbp})$, is
symmetric under the exchange of $\qp$ and $\qbp$. When we identify the primary
quark $q$, we find
\begin{eqnarray}
  \XTZ{E}{q}(q^2,z) &=& -4\mathbf{I}_{qg,F}^{(1)}(\e,q^2)\delta(1-z)
  + (q^2)^{-\e}\bigg[
    -\frac{5}{9} \delta(1-z)
    \nonumber \\ &&                         
    + \frac{1}{3} \Dn{0}(1-z)
    -\frac{1}{3}-\frac{1}{3}z+\frac{1}{6}z^2
    \bigg]   + \mathcal{O}(\e)\,;
  \label{eq:intE30q}
\end{eqnarray}
when we identify the secondary quark $\qp$ or $\qbp$, we get
\begin{eqnarray}
  \XTZ{E}{\qp}(q^2,z) &=& 
  (q^2)^{-\e}\bigg[ -\frac{1}{2\e}p^{(0)}_{qg}(z)
    + \log(z) \left(1-2z+2z^2\right)
    \nonumber \\ &&                             
    - \log(1-z) \left(-\frac{1}{2}+z-z^2\right)
    + \frac{3}{2}z -2z^2
    \bigg]   + \mathcal{O}(\e)\,.
  \label{eq:intE30qp}  
\end{eqnarray}
In~\eqref{eq:intE30q}, we note the presence of the $\mathbf{I}_{qg,F}^{(1)}$
infrared oparator, but the absence of a pole proportional to a splitting kernel,
since there are no collinear limits between quarks of different
flavour. Instead, when we identify a secondary quark, we do not have any
infrared operator, since the secondary quark flavour is absent at the virtual
level, but we have $p^{(0)}_{qg}$, encoding the splitting $g \to \qp \qbp$.

Finally, we have the gluon-gluon antenna function with all gluons
$F_3^0(1_g,2_g,3_g)$, whose integral reads
\begin{eqnarray}
  \XTZ{F}{g}(q^2,z) &=&
  -4\mathbf{I}^{(1)}_{gg}(\e,q^2)\delta(1-z)              
  + (q^2)^{-\e} \bigg[ -\frac{1}{\e}p^{(0)}_{gg}(z)
    + \delta(1-z)\left(\frac{67}{18} + \frac{2\pi^2}{3}\right)
    \nonumber \\ &&                    
    -\frac{11}{6} \Dn{0}(1-z)
    + 2\Dn{1}(1-z)
    + \log(z) \left(-8+\frac{4}{z}+\frac{4}{1-z} +4z-4z^2\right)
    \nonumber \\ &&                    
    -\log(1-z) \left(4-\frac{2}{z}-2z+2z^2\right)
    + \frac{11}{6} + \frac{11}{6}z+ \frac{11}{6}z^2
    \bigg]   + \mathcal{O}(\e)\,,
\end{eqnarray}
and the gluon-gluon antenna function with a quark pair
$G_3^0(1_g,2_{\qp},3_{\qbp})$, symmetric under exchange of the quark line: when
we identify the gluon, we obtain
\begin{eqnarray}
  \XTZ{G}{g}(q^2,z) &=& -2\mathbf{I}_{gg,F}^{(1)}(\e,q^2)\delta(1-z)
  + (q^2)^{-\e}\bigg[ - \frac{5}{9} \delta(1-z)
    \nonumber \\ &&             
    + \frac{1}{3}\Dn{0}(1-z) -\frac{1}{3} - \frac{1}{3}z - \frac{1}{3}z^2
    \bigg]   + \mathcal{O}(\e)\,,
\end{eqnarray}
whereas when we identify a quark we get
\begin{eqnarray}
  \XTZ{G}{q}(q^2,z) &=&
  (q^2)^{-\e}\bigg[-\frac{1}{2\e} p_{qg}^{(0)}(z)
    + \log(z) \left(1-2z+2z^2\right)
    \nonumber \\ &&                     
    - \log(1-z) \left( -\frac{1}{2} + z - z^2 \right)
    + z - \frac{7}{4} z^2
    \bigg]   + \mathcal{O}(\e)\,.
\end{eqnarray}

\bibliographystyle{JHEP}
\bibliography{FFfrag}

\end{document}